\documentclass[10pt]{article}
\usepackage{epsfig,citesort}
\usepackage{amssymb,amsmath}
\usepackage{lscape,graphicx}
\usepackage{setspace}

      \input{amssym.def}
      \input{amssym}
      \newcommand {\mm}[1] {\ifmmode{#1}\else{\mbox{\(#1\)}}\fi}

\DeclareMathOperator{\var}{var}

\begin{document}
 \title{\bf Interstrand pairing patterns in
  $\beta$-barrel membrane proteins: the positive-outside rule,
  aromatic rescue, and strand registration prediction}
 \author{\bf Ronald Jackups, Jr.
  and Jie Liang\thanks{Corresponding author. Phone:
  (312)355--1789, fax: (312)996--5921, email: {\tt jliang@uic.edu}} \\
  Department of Bioengineering, SEO, MC-063 \\ University of Illinois
  at Chicago\\ 851 S.\ Morgan Street, Room 218 \\ Chicago, IL
  60607--7052, U.S.A.
\\Journal reference: {\it  J. Mol. Biol. (2005) 354:979--993.}
}  \date{\today}

 \maketitle

\abstract{

$\beta$-barrel membrane proteins are found in the outer membrane of
gram-negative bacteria, mitochondria, and chloroplasts.  Little is
known about how residues in membrane $\beta$-barrels interact
preferentially with other residues on adjacent strands.  We have
developed probabilistic models to quantify propensities of residues
for different spatial locations and for interstrand pairwise contact
interactions involving strong H-bonds, side-chain interactions, and
weak H-bonds.  Using the reference state of exhaustive permutation of
residues within the same $\beta$-strand, the propensity values and
{\it p}-values measuring statistical significance are calculated
exactly by analytical formulae we have developed.  Our findings show
that there are characteristic preferences of residues for different
membrane locations.  Contrary to the ``positive-inside'' rule for
helical membrane proteins, $\beta$-barrel membrane proteins follow a
significant albeit weaker ``positive-outside'' rule, in that the basic
residues Arg and Lys are disproportionately favored in the
extracellular cap region and disfavored in the periplasmic cap region.
We find that different residue pairs prefer strong backbone H-bonded
interstrand pairings ({\it e.g.}\ Gly-Aromatic) or non-H-bonded
pairings ({\it e.g.}\ Aromatic-Aromatic).  In addition, we find that
Tyr and Phe participate in aromatic rescue by shielding Gly from polar
environments.  We also show that these propensities can be used to
predict the registration of strand pairs, an important task for the
structure prediction of $\beta$-barrel membrane proteins.  Our
accuracy of 44\% is considerably better than random (7\%).  It also
significantly outperforms a comparable registration prediction for
soluble $\beta$-sheets under similar conditions.  Our results imply
several experiments that can help to elucidate the mechanisms of {\it
in vitro\/} and {\it in vivo\/} folding of $\beta$-barrel membrane
proteins.  The propensity scales developed in this study will also be
useful for computational structure prediction and for folding
simulations.  

The most interesting part about the computational models is contained in the
supplementary information after the bibliography towards the end of the paper.
}
\vspace*{2cm}


\renewcommand{\topfraction}{1.0}

\section{Introduction}
Integral membrane proteins can be categorized into two structural
classes: $\alpha$-helical proteins and $\beta$-barrel proteins.
$\beta$-barrel proteins are found in the outer membrane of
gram-negative bacteria, mitochondria, and chloroplasts
\cite{Montoya03_BBA,Tamm01_JBC,Wimley03_COSB}.  It is estimated that
membrane $\beta$-barrels constitute about 2--3\% of all proteins in
the genomes of gram-negative bacteria \cite{Wimley03_COSB}.  Many
bacterial pore-forming exotoxins such as $\alpha$-hemolysin of \it
Staphylococcus aureus \rm and protective antigen of \it Bacillus
anthracis \rm are also $\beta$-barrel membrane proteins
\cite{Song96_Sci,Nassi02_BC}.  $\beta$-barrel proteins have diverse
biological functions, {\it e.g.}\ pore formation, membrane anchoring,
enzyme activity, and bacterial virulence \cite{Koebnik00_MM}.  Their
medical relevance is immediate, as membrane proteins in bacteria
provide candidate molecular targets for the development of
antimicrobial drugs and vaccines.

With recent rapid progress in the development of computational tools,
$\beta$-barrel membrane proteins can be reliably identified from
sequences \cite{Bigelow04_NAR,Martelli02_Bioinfo,Gromiha04_JCC}.
However, there are less than 30 structures currently known at atomic
details.  Based on knowledge of these structures, $\beta$-barrel
membrane proteins are known to contain an even number of strands (8 to
22).  These strands are arranged in an antiparallel $\beta$-sheet that
is twisted and tilted into a barrel structure, usually with short
$\beta$-turn-like periplasmic loops and long extracellular loops
\cite{Schulz00_COSB}.  A $\beta$-barrel protein may exist either as a
monomer or as an oligomer.  A set of simple rules has been compiled by
Schulz to describe the structural features of transmembrane
$\beta$-barrels \cite{Schulz00_COSB,Schulz02_BBA}.

Compared to helical membrane proteins, the nature of the assembly of
$\beta$-barrels is likely to be different, as there are many more
polar and ionizable residues intervening in the transmembrane (TM)
region.  Indeed, several studies have already identified significant
differences in folding between $\beta$-barrel and $\alpha$-helical
membrane proteins \cite{Tamm01_JBC,Tamm04_BBA,Montoya03_BBA}.  This
complexity makes the prediction of the TM strands of $\beta$-barrel
proteins from sequence more difficult than for helical proteins, as
simple rules based on stretches of hydrophobicity are no longer
effective \cite{Seshadri98_PS,Bigelow04_NAR,Gromiha04_JCC}.
Furthermore, because of the low sequence identity for TM strands in
$\beta$-barrel membrane proteins, comparative modeling cannot generate
reliable structures.

Despite these difficulties, recent studies have identified
characteristic amino acid preferences for different regions of the
$\beta$-barrel
\cite{Wimley02_PS,Gromiha04_JCC,Seshadri98_PS,Ulmschneider01_BBA,Chamberlain04_BJ}.
For example, polar residues are found to be enriched in the internal
face of the barrel and the solvent exposed regions of the protein
extending past the membrane bilayer.  Large aliphatic residues are
enriched in the external face of the barrel, and bulky aromatic
residues tend to prefer the external face at the headgroup regions,
forming ``girdles'' at the membrane interface
\cite{Schulz00_COSB,Wimley02_PS}.  Bigelow {\it et al}.\
also found that Tyr prefers the C-terminal over the N-terminal of a TM
strand \cite{Bigelow04_NAR}.

A number of important questions remain unanswered.  For example,
little is known about specific interactions between residues on
adjacent TM strands. In contrast, several studies have identified
various preferred two-residue interstrand interactions in soluble
$\beta$-sheets \cite{Wouters95_Prot,Steward02_Prot}.  A further
challenging question is whether the requirements of thermodynamic
stability in the folding and {\it in vivo\/} sorting, targeting, and
translocation of $\beta$-barrel membrane proteins lead to the
avoidance of certain types of spatial patterns.  Finally, we do not
know how residues in $\beta$-barrel membrane proteins interact with
specific regions of lipid bilayers.

To answer these important questions, we have developed methods to
estimate the positional propensity of residues for different spatial
regions of the barrel and to estimate the propensity for interstrand
pairwise spatial interactions.  Our findings show that there are
additional, previously unknown characteristic preferences of residues
for different locations. Contrary to the ``positive-inside'' rule for
$\alpha$-helical membrane proteins, $\beta$-barrel proteins follow the
``positive-outside'' rule, in that the basic residues Arg and Lys are
disproportionately favored in the extracellular cap region and
disfavored in the periplasmic cap region.  We also find several
notable interstrand pairwise motifs.  The Gly-Tyr and Gly-Phe motifs,
found in interstrand pairs sharing backbone H-bonds, reveal the
existence of ``aromatic rescue'' in $\beta$-barrel membrane proteins,
in which a large aromatic side-chain shields Gly from a polar
environment, a behavior first discovered in soluble $\beta$-sheets
\cite{Merkel98_FD}.  The propensity scale of interstrand pairwise
contact interactions we developed (called TransMembrane Strand
Interaction Propensity [{\sc Tmsip}]) is based on a physical model of
strand interactions involving backbone strong H-bond interactions,
non-H-bonded side-chain interactions, and weak H-bond interactions.
We also show that this propensity scale can be used to predict the
registration of strand pairs, an important task for the structure
prediction of $\beta$-barrel membrane proteins.  Our accuracy of 44\%
is better than random (7\%), and it significantly outperforms a
comparable registration prediction for soluble $\beta$-sheets under
similar conditions.  Our results also suggest several experimentally
testable hypotheses about the {\it in vitro\/} and {\it in vivo\/}
folding of $\beta$-barrel membrane proteins.

\section{Results}

Our study is based on a very small dataset of known structures,
comprising only 19 $\beta$-barrel membrane proteins.  It is a very
challenging task to identify significant patterns from so little data.
To analyze small sample data, our approach is to employ a rigorous
statistical model that enables the exact calculation of $p$-values
from observed patterns.  With the aid of a properly calculated
$p$-value, we can then identify patterns that are truly significant,
and discard spurious findings.  This is critical for spatial pattern
discovery in $\beta$-barrel membrane proteins, and ours is the first
study to employ such rigorous methods.

\subsection{Spatial regional preference of residues}
\subsubsection{Spatial regions of TM barrels }
\begin{figure}[t!]
      \centerline{\epsfig{figure=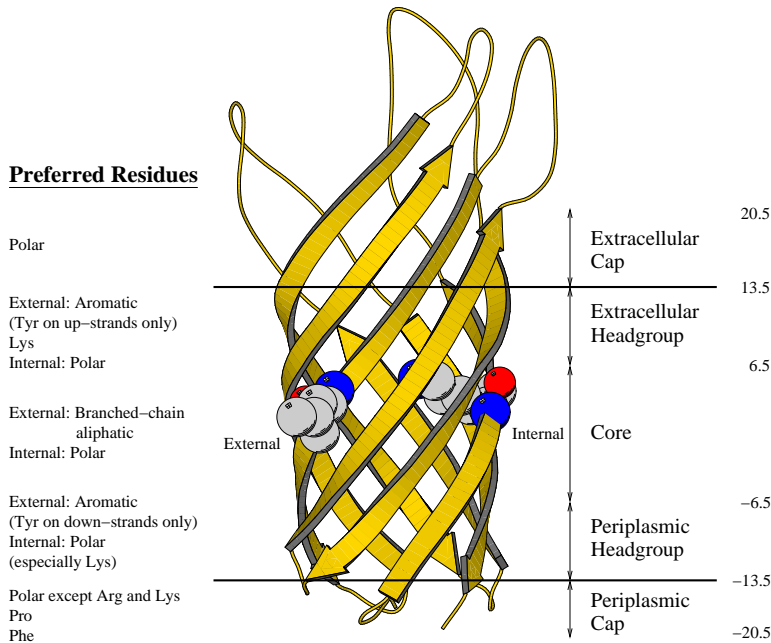,width=4in}}
\caption{\small \sf Definitions of boundaries of regions and summary
of residue region-specific preferences shown on the structure of the
8-strand barrel of OmpA.  Regions, defined by distance from the barrel
center, are defined as follows: Extracellular Cap = 13.5 \AA\ to 20.5 \AA\,
Extracellular Headgroup = 6.5 \AA\ to 13.5 \AA\, Core = $-6.5$ \AA\ to 6.5 \AA\,
Periplasmic Headgroup = $-13.5$ \AA\ to $-6.5$ \AA\, Periplasmic Cap = $-20.5$
\AA\ to $-13.5$ \AA\.  Preferred residues (those with high odds ratios) in each
region are listed on the left.  For reference, one internal residue
(Lys 12, on the right) and one external residue (Leu 79, on the left)
are shown.}
\label{Fig:cutoff}
\end{figure}

We define eight distinct spatial regions for each TM strand based on
the vertical distance along the membrane normal and the orientation of
the side-chain (inside-facing or outside-facing,
Figure~\ref{Fig:cutoff}).  After placing the origin of the reference
frame at the midpoint of the membrane bilayer and taking the vertical
axis perpendicular to the bilayer as the $z$-axis, we measure the
vertical distance of each residue from the barrel center (the $xy$
plane where $z=0$) along the $z$-axis.  We follow the definition of
Wimley \cite{Wimley02_PS} and define the {\it core region\/} as all
residues whose $\alpha$-carbons are within 6.5 \AA\ of the barrel
center.  The {\it periplasmic} and {\it extracellular headgroup
regions\/} are similarly defined for the space 6.5-13.5 \AA\ away from
the barrel center on either side of the core region. These two
headgroup regions are distinguished based on information in the
original paper of the protein structure.  Usually, both the N- and
C-termini of the peptide chain as well as the short loop regions are
found on the periplasmic side of the $\beta$-barrel.  The core and two
headgroup regions are further divided into {\it internal\/} and {\it
external\/} residues, depending on whether their side-chains face into
or away from the center of the barrel, respectively.  The {\it
periplasmic} and {\it extracellular cap regions\/} are defined as the
space 13.5-20.5 \AA\ away from the barrel center.  These cap regions
are similar to the membrane-water interface studied by Granseth {\it
et al.}\ \cite{Granseth05_JMB}.  Since the cap regions are not located
in the membrane domain, they are not divided into internal and
external regions.

\subsubsection{Region specific single-body propensities }

\renewcommand{\baselinestretch}{1.0}
\begin{table}[t!] 
\begin{small}
\begin{center}
       \vspace{4mm}
\begin{tabular}{ccl|cl|cl|cl}
      \hline \hline 
\multicolumn{1}{c}{} & \multicolumn{2}{c}{\scriptsize Periplasmic Cap} & \multicolumn{4}{c}{\scriptsize Periplasmic Headgroup} &
\multicolumn{2}{c}{\scriptsize Core} \\ \cline{2-9}
\multicolumn{1}{c}{\scriptsize Amino} & \multicolumn{2}{c}{} & \multicolumn{2}{c}{\scriptsize Internal} & \multicolumn{2}{c}{\scriptsize External} & \multicolumn{2}{c}{\scriptsize Internal} \\ \cline{2-9}
\scriptsize Acid & \scriptsize Odds & \multicolumn{1}{c}{\scriptsize $p$-Value} & \scriptsize Odds & \multicolumn{1}{c}{\scriptsize $p$-Value} & \scriptsize Odds & \multicolumn{1}{c}{\scriptsize $p$-Value} & \scriptsize Odds & \multicolumn{1}{c}{\scriptsize $p$-Value} \\
\hline
A & 0.72 & 2.4$\times 10^{-2}$  & 0.57 & 2.2$\times 10^{-2}$ & 0.61 & 3.1$\times 10^{-2}$  & 1.03 & \;\;\;\;\;\;--                  \\
R & 0.61 & 2.3$\times 10^{-2}$  & 1.62 & 3.4$\times 10^{-2}$ & 0.00 & 7.7$\times 10^{-7}$  & 1.77 & 3.2$\times 10^{-5}$  \\
N & 1.51 & 1.3$\times 10^{-3}$  & 1.45 & \;\;\;\;\;\;--      & 0.32 & 7.5$\times 10^{-4}$  & 1.01 & \;\;\;\;\;\;--      \\
D & 1.83 & 6.5$\times 10^{-8}$  & 1.33 & \;\;\;\;\;\;--      & 0.05 & 1.8$\times 10^{-8}$  & 0.73 & \;\;\;\;\;\;--      \\
C & --   & \;\;\;\;\;\;--       & --   & \;\;\;\;\;\;--      & --   & \;\;\;\;\;\;--       & --   & \;\;\;\;\;\;--      \\
Q & 0.52 & 4.6$\times 10^{-3}$  & 1.28 & \;\;\;\;\;\;--      & 0.82 & \;\;\;\;\;\;--       & 1.39 & 3.5$\times 10^{-2}$  \\
E & 1.20 & \;\;\;\;\;\;--       & 1.61 & 4.3$\times 10^{-2}$ & 0.00 & 1.4$\times 10^{-6}$  & 1.73 & 1.1$\times 10^{-4}$  \\
G & 1.20 & 4.7$\times 10^{-2}$  & 0.58 & 4.0$\times 10^{-3}$ & 0.67 & 2.1$\times 10^{-2}$  & 1.68 & 5.9$\times 10^{-11}$ \\
H & 0.59 & \;\;\;\;\;\;--       & 1.53 & \;\;\;\;\;\;--      & 2.61 & 4.9$\times 10^{-3}$  & 0.70 & \;\;\;\;\;\;--      \\
I & 0.74 & \;\;\;\;\;\;--       & 0.53 & \;\;\;\;\;\;--      & 1.90 & 3.5$\times 10^{-3}$  & 0.37 & 1.9$\times 10^{-4}$  \\
L & 0.86 & \;\;\;\;\;\;--       & 0.49 & 5.4$\times 10^{-3}$ & 1.65 & 1.0$\times 10^{-3}$  & 0.42 & 1.9$\times 10^{-7}$  \\
K & 0.86 & \;\;\;\;\;\;--       & 2.37 & 2.3$\times 10^{-5}$ & 0.16 & 3.2$\times 10^{-4}$  & 1.16 & \;\;\;\;\;\;--      \\
M & 0.89 & \;\;\;\;\;\;--       & 0.38 & \;\;\;\;\;\;--      & 0.53 & \;\;\;\;\;\;--       & 0.88 & \;\;\;\;\;\;--      \\
F & 1.41 & 1.6$\times 10^{-2}$  & 0.13 & 3.1$\times 10^{-5}$ & 2.70 & 6.8$\times 10^{-10}$ & 0.39 & 2.7$\times 10^{-5}$  \\
P & 2.79 & 1.2$\times 10^{-10}$ & 1.03 & \;\;\;\;\;\;--      & 0.27 & 3.7$\times 10^{-2}$  & 0.07 & 2.8$\times 10^{-6}$  \\
S & 0.94 & \;\;\;\;\;\;--       & 1.78 & 5.0$\times 10^{-4}$ & 0.39 & 8.3$\times 10^{-4}$  & 1.33 & 1.6$\times 10^{-2}$  \\
T & 0.98 & \;\;\;\;\;\;--       & 1.80 & 3.4$\times 10^{-4}$ & 0.77 & \;\;\;\;\;\;--       & 1.14 & \;\;\;\;\;\;--      \\
W & 0.82 & \;\;\;\;\;\;--       & 0.41 & \;\;\;\;\;\;--      & 2.54 & 1.5$\times 10^{-4}$  & 0.50 & 2.8$\times 10^{-2}$  \\
Y & 0.24 & 2.1$\times 10^{-10}$ & 0.51 & 1.7$\times 10^{-2}$ & 2.46 & 5.2$\times 10^{-10}$ & 0.95 & \;\;\;\;\;\;--      \\
V & 0.65 & 1.2$\times 10^{-2}$  & 0.38 & 2.3$\times 10^{-3}$ & 1.85 & 2.6$\times 10^{-4}$  & 0.30 & 2.0$\times 10^{-8}$  \\
      \hline \hline 
\multicolumn{1}{c}{} & \multicolumn{2}{c}{\scriptsize Core} & \multicolumn{4}{c}{\scriptsize Extracellular Headgroup} &
\multicolumn{2}{c}{\scriptsize Extracellular Cap} \\ \cline{2-9}
\multicolumn{1}{c}{\scriptsize Amino} & \multicolumn{2}{c}{\scriptsize External} & \multicolumn{2}{c}{\scriptsize Internal} & \multicolumn{2}{c}{\scriptsize External} & \multicolumn{2}{c}{} \\ \cline{2-9}
\scriptsize Acid & \scriptsize Odds & \multicolumn{1}{c}{\scriptsize $p$-Value} & \scriptsize Odds & \multicolumn{1}{c}{\scriptsize $p$-Value} & \scriptsize Odds & \multicolumn{1}{c}{\scriptsize $p$-Value} & \scriptsize Odds & \multicolumn{1}{c}{\scriptsize $p$-Value} \\
\hline

A & 1.95 & 2.8$\times 10^{-13}$ & 0.81 & \;\;\;\;\;\;--       & 0.42 & 1.2$\times 10^{-3}$  & 1.00 & \;\;\;\;\;\;--     \\
R & 0.07 & 1.0$\times 10^{-11}$ & 1.15 & \;\;\;\;\;\;--       & 0.87 & \;\;\;\;\;\;--       & 1.54 & 2.2$\times 10^{-4}$ \\
N & 0.13 & 4.7$\times 10^{-13}$ & 1.03 & \;\;\;\;\;\;--       & 0.36 & 2.7$\times 10^{-3}$  & 1.56 & 7.0$\times 10^{-6}$ \\
D & 0.09 & 1.2$\times 10^{-15}$ & 1.11 & \;\;\;\;\;\;--       & 0.76 & \;\;\;\;\;\;--       & 1.53 & 6.5$\times 10^{-6}$ \\
C & --   & \;\;\;\;\;\;--       & --   & \;\;\;\;\;\;--       & --   & \;\;\;\;\;\;--       & 4.41 & \;\;\;\;\;\;--     \\
Q & 0.14 & 3.1$\times 10^{-9}$  & 2.22 & 7.8$\times 10^{-5}$  & 1.23 & \;\;\;\;\;\;--       & 1.18 & \;\;\;\;\;\;--     \\
E & 0.00 & 5.4$\times 10^{-14}$ & 1.28 & \;\;\;\;\;\;--       & 0.41 & 2.9$\times 10^{-2}$  & 1.32 & 3.3$\times 10^{-2}$ \\
G & 0.75 & 7.2$\times 10^{-3}$  & 1.15 & \;\;\;\;\;\;--       & 0.39 & 1.7$\times 10^{-5}$  & 0.96 & \;\;\;\;\;\;--     \\
H & 0.23 & 8.2$\times 10^{-3}$  & 0.25 & \;\;\;\;\;\;--       & 2.61 & 8.1$\times 10^{-3}$  & 1.06 & \;\;\;\;\;\;--     \\
I & 2.29 & 1.7$\times 10^{-10}$ & 0.79 & \;\;\;\;\;\;--       & 1.00 & \;\;\;\;\;\;--       & 0.60 & 5.2$\times 10^{-3}$ \\
L & 2.49 & 9.7$\times 10^{-29}$ & 0.65 & \;\;\;\;\;\;--       & 0.89 & \;\;\;\;\;\;--       & 0.51 & 1.4$\times 10^{-7}$ \\
K & 0.04 & 8.3$\times 10^{-12}$ & 0.75 & \;\;\;\;\;\;--       & 1.21 & \;\;\;\;\;\;--       & 1.54 & 3.6$\times 10^{-4}$ \\
M & 1.53 & \;\;\;\;\;\;--       & 1.51 & \;\;\;\;\;\;--       & 0.98 & \;\;\;\;\;\;--       & 0.98 & \;\;\;\;\;\;--     \\
F & 1.26 & \;\;\;\;\;\;--       & 0.26 & 7.5$\times 10^{-4}$  & 1.89 & 1.3$\times 10^{-3}$  & 0.61 & 1.5$\times 10^{-3}$ \\
P & 1.11 & \;\;\;\;\;\;--       & 0.00 & 1.4$\times 10^{-3}$  & 0.60 & \;\;\;\;\;\;--       & 1.03 & \;\;\;\;\;\;--     \\
S & 0.29 & 5.1$\times 10^{-10}$ & 1.48 & 2.7$\times 10^{-2}$  & 0.43 & 3.5$\times 10^{-3}$  & 1.30 & 5.2$\times 10^{-3}$ \\
T & 0.70 & 1.6$\times 10^{-2}$  & 1.55 & 1.2$\times 10^{-2}$  & 0.75 & \;\;\;\;\;\;--       & 0.86 & \;\;\;\;\;\;--     \\
W & 0.85 & \;\;\;\;\;\;--       & 0.54 & \;\;\;\;\;\;--       & 3.63 & 6.6$\times 10^{-9}$  & 0.57 & 1.8$\times 10^{-2}$ \\
Y & 1.11 & \;\;\;\;\;\;--       & 0.75 & \;\;\;\;\;\;--       & 2.91 & 6.2$\times 10^{-14}$ & 0.62 & 3.1$\times 10^{-4}$ \\
V & 2.65 & 4.9$\times 10^{-26}$ & 0.69 & \;\;\;\;\;\;--       & 1.10 & \;\;\;\;\;\;--       & 0.52 & 6.0$\times 10^{-6}$ \\
      \hline \hline 
\end{tabular}
\end{center}
\end{small}
      \caption{\small Single-body propensities by region with significance
level measured by $p$-values.  Odds ratio entries listed as ``--''
indicate that no residues of that type occurred in the region in the
dataset.  $p$-value entries listed as ``--'' are not significant at
the threshold of 0.05.}  \label{tab:fullsingle}
\end{table}

Single-body propensities can be used to reflect the preference of
different types of residues for different spatial regions of the
transmembrane (TM) barrel.  We define the single-body propensity of an
amino acid residue type as an odds ratio comparing the frequency of
this residue type in one region to its frequency in all eight regions
combined.  Results of residue single-body propensities are listed in
Table~\ref{tab:fullsingle}, along with statistical significance in the
form of $p$-values.  These two-tailed $p$-values represent the
probability that an odds-ratio will be more extreme than the observed
one in the reference model, in which the positions of the residues in
all proteins in the dataset are equally likely to take any one
combination when they are exhaustively permuted.

\subsubsection{Location preference of residues}

Similar to previous studies \cite{Wimley02_PS}, we find that aromatic
residues have a strong preference for the external extracellular and
periplasmic headgroup regions.  Long aliphatic residues are favored in
the external-facing core regions, and polar residues in all
internal-facing regions as well as both cap regions.  Cys residues do
not occur in any TM region or the periplasmic cap region, consistent
with earlier findings for helical membrane proteins
\cite{Adamian01_JMB}.  The favorable propensity of Cys for the
extracellular cap in Table~\ref{tab:fullsingle} is due to a single
cysteine cross-bridge in LamB (pdb {\tt 2mpr}) and is not
statistically significant.

We find that Pro prefers to be in the periplasmic cap region, because
Pro residues are enriched in the short $\beta$-turns found in this
region.  Phe also prefers this region, despite its generally low
propensity for $\beta$-turn regions \cite{Wilmot88_JMB}.  While the
$\alpha$-carbon of the Phe residue is situated in the periplasmic cap
and therefore does not contribute to the TM $\beta$-sheet, the
aromatic side-chain extends into the TM region and contributes to the
``aromatic girdle'' by ``anti-snorkeling.''  The preference of Phe for
the periplasmic cap region has been described for both
$\alpha$-helical \cite{Chamberlain04_JMB} and $\beta$-barrel membrane
proteins \cite{Schulz04_BAEP}.

Our study provides novel insights about the location preferences of
residues, as well as the significance of these preferences as measured
by $p$-values (Table~\ref{tab:fullsingle}), which have not been
provided in previous studies \cite{Wimley02_PS,Gromiha03_IJBM}.  We
discuss several novel findings in more detail below.

\paragraph{The positive-outside rule: asymmetric distribution of basic residues in the cap regions.\/}
The propensities for basic amino acids are very different between the
periplasmic and extracellular cap regions.  Arg and Lys have a lower
preference for the periplasmic cap region (0.61 and 0.86 respectively)
than for the extracellular cap regions (1.54 for both), despite the
general ``rule'' that polar residues prefer both cap regions.  Thus
basic residues are more than twice as likely to be found in the
extracellular cap than in the periplasmic cap.  In contrast, acidic
residues (Asp and Glu) are prevalent and evenly distributed in both
cap regions.  This pattern is analogous to the ``positive-inside''
rule for $\alpha$-helical membrane proteins, which states that acidic
residues are evenly distributed at both ends of a TM helix, but basic
residues favor the {\it cis} side of the membrane, {\it i.e.}\ the
side in which the helix inserts \cite{vonHeijne89_Nat}.  Because
$\beta$-barrel membrane proteins insert into the outer membrane from
the periplasmic side \cite{Tamm04_BBA}, the asymmetric distribution of
basic residues described here is the {\it opposite} of the
positive-inside rule, and thus we have named it the
``positive-outside'' rule.

\subsection{Strand interactions: pairwise spatial motifs and antimotifs}
Single-residue propensities cannot capture information about
interactions between residues on different strands.  Little is known
about whether there is any specific preference for residues to
interact across adjacent strands.  This is in contrast to helical
membrane proteins, where helical interactions are the subject of
several studies
\cite{Adamian01_JMB,Adamian02_Prot,Adamian03_JMB,Eilers02_BJ,Liu04_JMB,Gimpelev04_BJ}.

\subsubsection{Model of strand interactions: Strong H-bonds, side-chain interactions, and weak H-bonds}

\begin{figure}[t!]
      \centerline{\epsfig{figure=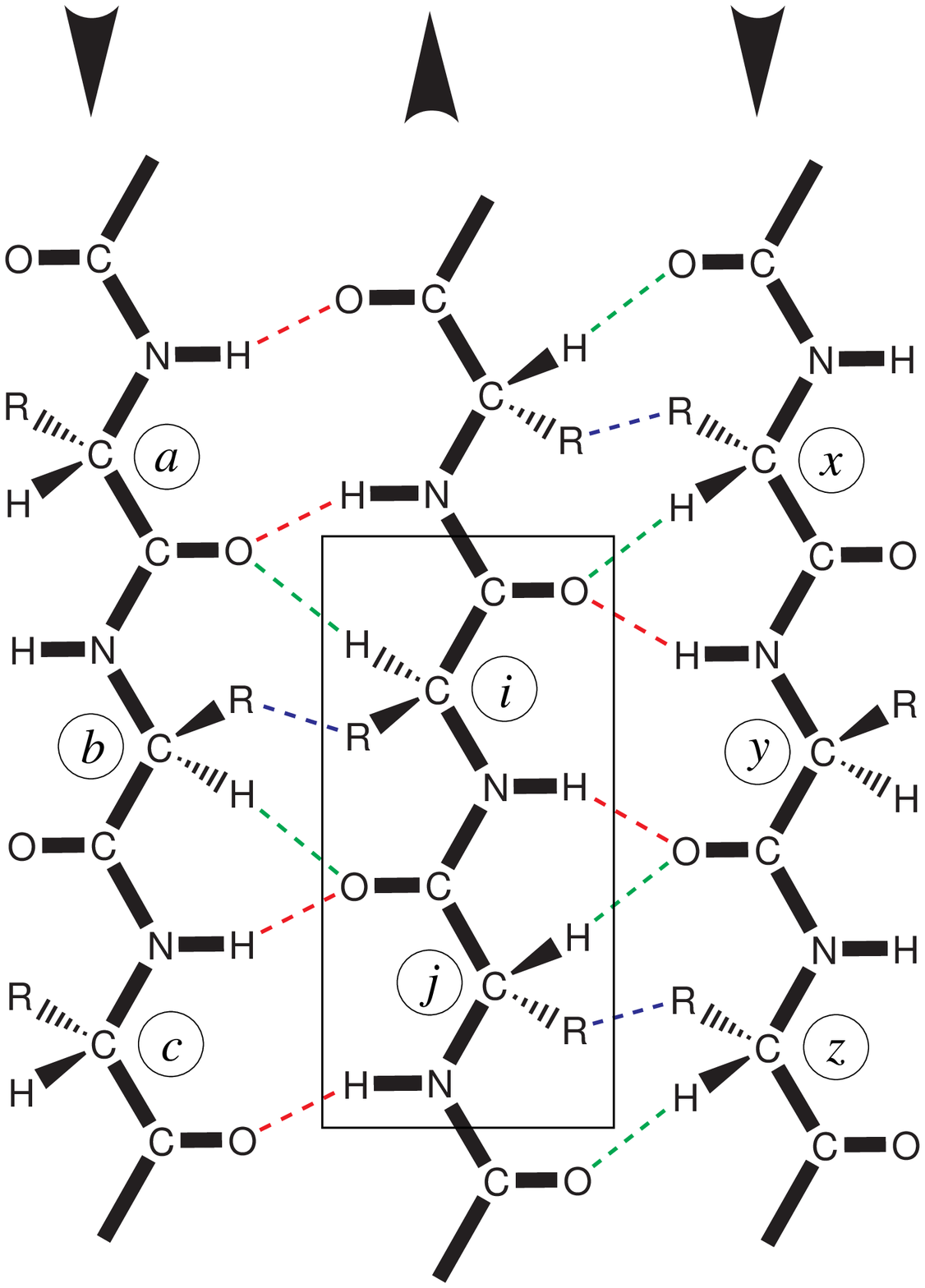,width=3.5in}}
\caption{\small \sf Repeating pattern of interstrand interactions and
alternating internal and external side-chains.  Red = strong H-bonds,
Blue = non-H-bonded interactions, Green = weak H-bonds.  Residues $i$
and $j$ in the center constitute one dyad (boxed).  Residue $i$ interacts in a
strong N-O H-bond with residue $y$ on its right and a non-H-bonded
interaction with residue $b$ on its left.  In contrast, residue $j$
interacts in a strong H-bond with residue $c$ on its {\it left} and a
non-H-bonded interaction with residue $z$ on its {\it right}.  This
alternating dyad pattern repeats throughout each strand.  In addition,
each residue also interacts through weak C$_\alpha$-O H-bonds to two
residues, one on each side.  In this case, $i$ shares a weak H-bond
with both $a$ and $x$, and $j$ with both $b$ and $y$.  Thus, residue
$j$ interacts through weak H-bonds to the ``bridge partners'' (strong
H-bond and non-H-bonded interactions, in this case $b$ and $y$) of the
next residue in the strand, $i$, in the N-C direction (as shown by arrows at the top).  }
\label{Fig:interaction}
\end{figure}

Strands in $\beta$-barrel membrane proteins have 7 or more residues
with a clear internal face and external face.  Interstrand
interactions in antiparallel $\beta$-sheets are characterized by a
periodic dyad bonding repeat pattern, which consists of units of two
residues (Figure~\ref{Fig:interaction}).  We characterize { \it
interstrand pairwise interactions\/} with three different types of
interactions, namely, strong regular hydrogen bonds between backbone N
and O atoms, ``non-H-bonded'' side-chain interactions (interactions
without strong backbone N-O hydrogen bonds), and weak C$_\alpha$-O
hydrogen bonds between the C$_\alpha$ atom of one residue and a
backbone oxygen on another strand.  These three interstrand
interactions occur periodically and are depicted in
Figure~\ref{Fig:interaction}.

The strong N-O hydrogen bonds occur between residues on adjacent
strands.  The non-H-bonded interactions alternate with the strong
hydrogen bonds along adjacent $\beta$-strands.  The weak C$_\alpha$-O
hydrogen bonds are displaced one residue along adjacent
$\beta$-strands.  Recent studies by Ho and Curmi on soluble
$\beta$-sheets have shown that these weak hydrogen bonds have a
significant impact on the arrangement of residues on adjacent strands,
such that bonded residues are not immediately across from each other,
but slightly displaced along the axis of the strand \cite{Ho02_JMB}.
This effect may be responsible for the twisting of $\beta$-sheets.
The role of weak C$_\alpha$-O H-bonds has also been well studied in
helical membrane proteins \cite{Senes01_PNAS}.

The three types of interactions used in this study are defined by
their relationship to the {\it backbone} hydrogen-bonding pattern in
$\beta$-sheets.  Unless otherwise specified, we use ``strong H-bond''
to refer to a pair of residues that share strong N-O backbone H-bonds,
``non-H-bonded interaction'' to a pair of residues on adjacent strands
that interact but do not share backbone H-bonds, and ``weak H-bond''
to a pair of residues that share weak C-O backbone H-bonds
(Figure~\ref{Fig:interaction}).

\subsubsection{Interstrand pairwise interaction propensity, spatial motifs and antimotifs}
We study interstrand interactions to identify spatial patterns in the
form of pairing residues across strands with a strong preference to
form hydrogen bonds, non-H-bonded interactions, and weak C$_\alpha$-O
hydrogen bonds.  We calculate TransMembrane Strand Interaction
Propensity ({\sc Tmsip}) scales, which are the values of odds ratios
comparing the observed frequency of interstrand contact interactions
to the expected frequency.  Among these contact interactions, we
define {\it spatial motifs\/} as two-body interactions whose
propensity is greater than 1.2 ({\it i.e.}\ favored) and statistically
significant, and {\it spatial antimotifs\/} as two-body interactions
whose propensity is less than 0.8 ({\it i.e.}\ disfavored) and
statistically significant.  Table~\ref{tab:interSig} lists the {\sc
Tmsip} values and the $p$-values of motifs and antimotifs for each
interaction type (strong H-bonds, non-H-bonded interactions, and weak
H-bonds) at the significance level of $p < 0.05$.  A full list of all
residue pairs is included in Supplementary Material.

\renewcommand{\baselinestretch}{1.0}
\begin{table}[t!] 
\setlength{\tabcolsep}{4pt}
\begin{center}
      \vspace{4mm} 
\begin{tabular}{crrccccrrccccrrcc}
      \hline \hline 
\multicolumn{17}{c}{\large (a) Motifs} \\
\multicolumn{5}{c}{\small Strong H-Bonds} & & \multicolumn{5}{c}{\small Non-H-Bonded Interactions} & & \multicolumn{5}{c}{\small Weak H-Bonds} \\
\cline{1-5} \cline{7-11} \cline{13-17}
\scriptsize Pair & \scriptsize Obs. & \scriptsize Exp. & \scriptsize Odds & \scriptsize $p$-Value & & \scriptsize Pair & \scriptsize Obs. & \scriptsize Exp. & \scriptsize Odds & \scriptsize $p$-Value & & \scriptsize Pair & \scriptsize Obs. & \scriptsize Exp. & \scriptsize Odds & \scriptsize $p$-Value \\
\cline{1-5} \cline{7-11} \cline{13-17}

GY & 39 & 25.0 & 1.56 & 8.0$\times 10^{-4}$ & & WY & 22 &  8.1 & 2.71 & 4.1$\times 10^{-7}$ & & DT & 17 &  8.3 & 2.05 & 2.2$\times 10^{-3}$ \\
ND & 10 &  3.6 & 2.76 & 1.2$\times 10^{-3}$ & & GI & 16 &  9.0 & 1.77 & 1.2$\times 10^{-2}$ & & GP & 12 &  6.8 & 1.78 & 1.3$\times 10^{-2}$ \\
GF & 18 & 10.0 & 1.80 & 4.7$\times 10^{-3}$ & & RE & 12 &  6.4 & 1.87 & 1.3$\times 10^{-2}$ & & EM &  6 &  2.6 & 2.31 & 3.5$\times 10^{-2}$ \\
IY & 18 & 10.1 & 1.79 & 4.8$\times 10^{-3}$ & & GV & 22 & 13.7 & 1.60 & 1.6$\times 10^{-2}$ & & DP &  5 &  2.0 & 2.52 & 3.9$\times 10^{-2}$ \\
KS & 12 &  6.2 & 1.95 & 1.2$\times 10^{-2}$ & & QG & 19 & 12.1 & 1.57 & 2.3$\times 10^{-2}$ & & & & & & \\
LW & 10 &  5.2 & 1.92 & 2.5$\times 10^{-2}$ & & LL & 24 & 16.7 & 1.44 & 2.7$\times 10^{-2}$ & & & & & & \\
LY & 34 & 24.7 & 1.37 & 2.9$\times 10^{-2}$ & & AV & 29 & 20.9 & 1.39 & 3.7$\times 10^{-2}$ & & & & & & \\
RP &  3 &  0.8 & 4.00 & 3.1$\times 10^{-2}$ & & LP &  8 &  3.9 & 2.07 & 3.7$\times 10^{-2}$ & & & & & & \\
AA & 16 & 10.6 & 1.50 & 4.8$\times 10^{-2}$ & & & & & & & & & & & & \\
HK &  3 &  0.9 & 3.33 & 5.0$\times 10^{-2}$ & & & & & & & & & & & & \\
      \hline \hline
\multicolumn{17}{c}{\large (b) Antimotifs} \\
\multicolumn{5}{c}{\small Strong H-Bonds} & & \multicolumn{5}{c}{\small Non-H-Bonded Interactions} & & \multicolumn{5}{c}{\small Weak H-Bonds} \\
\cline{1-5} \cline{7-11} \cline{13-17}
\scriptsize Pair & \scriptsize Obs. & \scriptsize Exp. & \scriptsize Odds & \scriptsize $p$-Value & & \scriptsize Pair & \scriptsize Obs. & \scriptsize Exp. & \scriptsize Odds & \scriptsize $p$-Value & & \scriptsize Pair & \scriptsize Obs. & \scriptsize Exp. & \scriptsize Odds & \scriptsize $p$-Value \\
\cline{1-5} \cline{7-11} \cline{13-17}

YY & 3 & 10.6 & 0.28 & 3.7$\times 10^{-3}$ & & GK &  3 &  9.4 & 0.32 & 1.0$\times 10^{-2}$ & & PT &  0 &  3.7 & 0.00 & 1.6$\times 10^{-2}$ \\
   &   &      &      &                     & & QV &  0 &  4.2 & 0.00 & 1.7$\times 10^{-2}$ & & AT & 22 & 33.2 & 0.66 & 1.9$\times 10^{-2}$ \\
   &   &      &      &                     & & GY & 14 & 22.7 & 0.62 & 2.7$\times 10^{-2}$ & & FV &  1 &  5.3 & 0.19 & 2.6$\times 10^{-2}$ \\
   &   &      &      &                     & & NL &  2 &  6.1 & 0.33 & 4.6$\times 10^{-2}$ & & & & & & \\
      \hline \hline

\end{tabular}
\end{center}
      \caption{\small Pairwise interstrand spatial motifs and antimotifs with propensities and $p$-values by
interaction type.  Only motifs significant at the threshold $p$-value of 0.05 are listed.  Obs. = observed
number of pairs, Exp. = expected number of pairs calculated as described in methods.} 
      \label{tab:interSig} 
\end{table}

For strong backbone H-bond interactions, small residues and aromatic
residues often have a strong propensity to interact, such as G-Y and
G-F.  Polar-polar residue interactions are also strongly favored,
including N-D and K-S.  In addition, branched hydrophobic residues and
aromatic residues frequently form spatial motifs (I-Y, L-W, and L-Y).
In contrast, aromatic-aromatic interactions are strongly disfavored
for strong H-bond interactions: Y-Y is an antimotif, and W-Y is also
strongly disfavored (propensity 0.21, $p$-value 0.06).  This likely
reflects the effects of bulky aromatic side-chains.

For non-backbone-H-bonded interactions, the W-Y motif stands out as
the one with the most significant $p$-value among all spatial motifs
(propensity 2.71, $p$-value 4$\times 10^{-7}$).  Small residues form
spatial motifs with branched residues and Gln (G-I, G-V, Q-G, and
A-V).  Other motifs of non-H-bonded interactions involve Leu (L-L and
L-P).

Weak H-bond interactions have fewer spatial motifs.  These involve
charged residues (D-T, E-M, and D-P), or Pro residues (G-P
and D-P).

\subsubsection{Propensity and spatial motifs using a reduced alphabet}
As seen in Table~\ref{tab:interSig}, residues of similar shape or
physicochemical properties often behave similarly in forming high
propensity spatial interactions.  It is known that a reduced alphabet
of fewer than 20 types of amino acid residues is often adequate for a
protein to perform its biological functions \cite{Riddle97_NSB}, for
computational recognition of homologous protein sequences
\cite{Murphy00_PE}, and for recognition of native structures from
decoys \cite{Li03_Prot}.  A reduced alphabet would also be useful for
statistical studies using limited data, as it would increase
statistical significance for groups of related amino acids.

\begin{figure}[t!]
      \centerline{\epsfig{figure=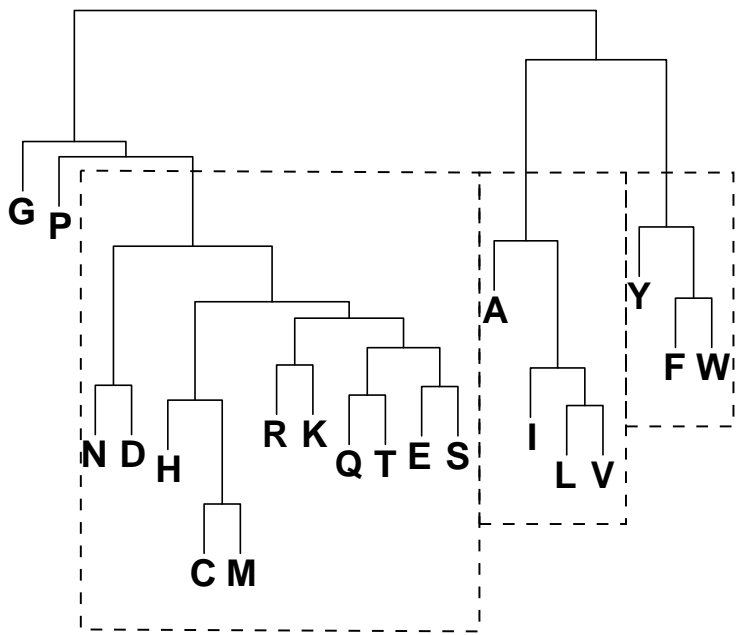,width=4in}}
\caption{\small \sf Clustering of residues based on single and pairwise
propensities.  G and P are assigned to their own clusters.  The remaining
three clusters are Polar (N, D, H, C, M, R, K, Q, T, E, and S), Aliphatic
(A, I, L, and V), and Aromatic (Y, F, and W). }
\label{Fig:cluster}
\end{figure}

To further understand the nature of interstrand propensity and spatial
motifs, we recalculated interstrand propensity using a simplified
amino acid alphabet.  An objective way to obtain a reduced alphabet is
to group amino acids with similar location preference and strand
pairing propensities.  We cluster amino acids using the results
obtained in the single and two-body studies, as described in
Supplementary Material.  The resulting clusters forming the reduced
alphabet (Figure~\ref{Fig:cluster}) are as follows: Gly by itself, Pro
by itself, polar residues (Arg, Asn, Asp, Cys, Gln, Glu, His, Lys,
Met, Ser, and Thr), aliphatic residues (Ala, Ile, Leu, and Val), and
aromatic residues (Phe, Trp, and Tyr).  Odds ratios and $p$-values
were then obtained for strand interactions between residue clusters
using the same method for the full amino acid alphabet.
Table~\ref{tab:interWeight} summarizes the results obtained using this
reduced alphabet.

\renewcommand{\baselinestretch}{1.0}
\begin{table}[t!] 
\setlength{\tabcolsep}{4pt}
\begin{center}
      \vspace{4mm} 
\begin{tabular}{lcrrcccrrcccrrcc}
      \hline \hline 
& & \multicolumn{4}{c}{\small Strong H-Bonds} & & \multicolumn{4}{c}{\small Non-H-Bonded} & & \multicolumn{4}{c}{\small Weak H-Bonds} \\
\cline{3-6} \cline{8-11} \cline{13-16}
\small Pair & & \scriptsize Obs. & \scriptsize Exp. & \scriptsize Odds & \scriptsize $p$-Value & & \scriptsize Obs. & \scriptsize Exp. & \scriptsize Odds & \scriptsize $p$-Value & & \scriptsize Obs. & \scriptsize Exp. & \scriptsize Odds & \scriptsize $p$-Value \\
\cline{1-1} \cline{3-6} \cline{8-11} \cline{13-16}

Gly-Gly & &        16 &  19.7 & 0.81 & --                  & &  14 &  17.8 & 0.78 & --                  & &  34 &  34.4 & 0.99 & -- \\
Gly-Pro & &         3 &   1.1 & 2.77 & --                  & &   3 &   2.9 & 1.05 & --                  & &  12 &   6.8 & 1.78 & 1.3$\times 10^{-2}$ \\
Gly-Polar & &     100 & 115.0 & 0.87 & 2.7$\times 10^{-2}$ & & 122 & 118.5 & 1.03 & --                  & & 142 & 161.4 & 0.88 & 2.0$\times 10^{-2}$ \\
Gly-Aliph. & &     68 &  68.6 & 0.99 & --                  & &  87 &  68.1 & 1.28 & 2.1$\times 10^{-3}$ & & 230 & 208.5 & 1.10 & 1.8$\times 10^{-2}$ \\
Gly-Arom. & &      62 &  40.9 & 1.52 & 2.5$\times 10^{-5}$ & &  26 &  40.8 & 0.64 & 2.6$\times 10^{-3}$ & &  84 &  90.6 & 0.93 & -- \\
Pro-Pro & &         0 &   0.0 & --   & --                  & &   0 &   0.6 & 0.00 & --                  & &   0 &   0.0 & --   & -- \\
Pro-Polar & &       8 &   4.6 & 1.74 & --                  & &   7 &   7.4 & 0.95 & --                  & &  15 &  18.3 & 0.82 & -- \\
Pro-Aliph. & &      3 &   8.6 & 0.35 & 5.3$\times 10^{-3}$ & &  15 &  10.1 & 1.49 & --                  & &   8 &   9.3 & 0.86 & -- \\
Pro-Arom. & &       4 &   3.7 & 1.09 & --                  & &   4 &   7.4 & 0.54 & --                  & &   3 &   3.6 & 0.83 & -- \\
Polar-Polar & &   228 & 204.9 & 1.11 & 1.1$\times 10^{-4}$ & & 239 & 210.5 & 1.14 & 8.6$\times 10^{-6}$ & & 223 & 216.5 & 1.03 & -- \\
Polar-Aliph. & &  167 & 178.2 & 0.94 & --                  & & 122 & 182.3 & 0.67 & 2.1$\times 10^{-11}$ & & 662 & 677.1 & 0.98 & -- \\
Polar-Arom. & &    67 &  90.3 & 0.74 & 2.6$\times 10^{-4}$ & &  99 &  98.7 & 1.00 & --                  & & 344 & 319.1 & 1.08 & 1.5$\times 10^{-2}$ \\ 
Aliph.-Aliph. & & 140 & 144.7 & 0.97 & --                  & & 182 & 152.5 & 1.19 & 3.4$\times 10^{-6}$ & & 166 & 162.6 & 1.02 & -- \\
Aliph.-Arom. & &  158 & 131.1 & 1.20 & 5.1$\times 10^{-4}$ & & 119 & 141.5 & 0.84 & 6.1$\times 10^{-3}$ & & 155 & 164.8 & 0.94 & -- \\
Arom.-Arom. & &    23 &  35.5 & 0.65 & 2.3$\times 10^{-3}$ & &  62 &  41.8 & 1.48 & 1.4$\times 10^{-5}$ & &  34 &  37.9 & 0.90 & -- \\

      \hline \hline                                          

\end{tabular}
\end{center}
      \caption{\small Odds ratios for interstrand residue pairs using
a reduced amino acid alphabet.  Residue clusters forming the reduced
alphabet are defined in Figure~\ref{Fig:cluster}.  $p$-value entries
listed as ``--'' were not significant at the threshold of 0.05.
Aliph. = Aliphatic, Arom. = Aromatic, Obs. = observed number of pairs,
Exp. = expected number of pairs calculated as described in methods.}
\label{tab:interWeight}
\end{table}

Several general patterns emerge.  The most dramatic differences in
strand pairing between different types of interactions are those
between strong H-bonds and non-H-bonded interactions.  Gly-Aromatic
and Aliphatic-Aromatic pairs have much higher preferences for strong
H-bonds than for non-H-bonded interactions, while Aromatic-Aromatic
and Gly-Aliphatic pairs have higher preferences for non-H-bonded
interactions.  Smaller differences are detected for Polar-Aliphatic
pairs (higher in strong H-bonds) and for Polar-Aromatic and
Aliphatic-Aliphatic pairs (higher in non-H-bonded interactions).  Weak
H-bond interactions did not show very strong preferences in this
analysis, except for the G-P pair already reported in
Table~\ref{tab:interSig}.

\begin{figure}[t!]
      \centerline{\epsfig{figure=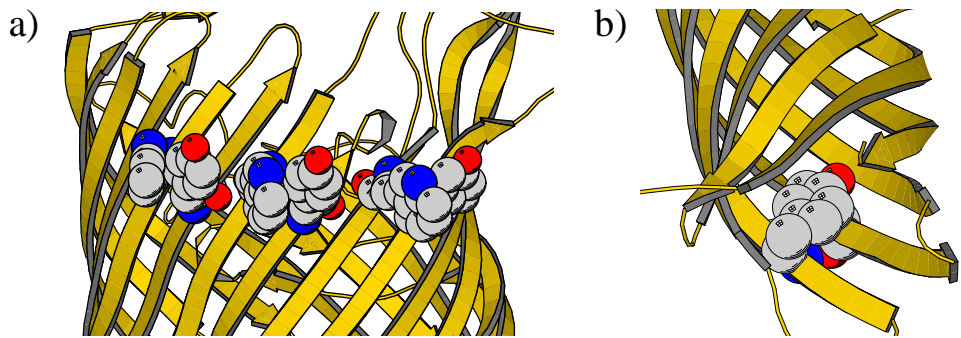,width=5in}}
\caption{\small Two examples of interstrand spatial motifs in
$\beta$-barrel membrane proteins: a) An instance of the WY
non-H-bonded interaction motif in LamB.  The aromatic
side-chains of Trp and Tyr show considerable contact interaction.
b) An instance of the GY strong H-bonded interaction
motif in NspA.  The protein has been tilted to show the motif on the internal
side of the barrel.  The aromatic side-chain of Tyr interacts with the Gly
residue on the adjacent strand.  This is an example of ``aromatic rescue.''}
\label{Fig:inter}
\end{figure}

The high-propensity Aromatic-Aromatic non-H-bonded interactions
include W-Y, the highest-propensity pairwise spatial motif.  An
example of the W-Y motif (Figure~\ref{Fig:inter}a) may help to explain
why this family of interactions has such high propensity.  The Tyr and
Trp residues involved in this motif often align their side-chains to
maximize contact between their aromatic rings and between their polar
groups.  At the same time, the bulky side-chains cause steric
hindrance preventing direct backbone N-O hydrogen bonds, and thus
these pairs are disfavored in strong backbone H-bond interactions.

\subsubsection{Internal and external preference of pairwise interactions}
We find that there are clear preferences for some residue pairs to
face either the outside or inside of the barrel.  Pairs of bulky
aromatic residues (Phe, Tyr, and Trp) are almost always on the outside
of the barrel, due to the steric hindrance in the barrel interior that
would result.  All 26 of the strong H-bonded Aromatic-Aromatic
interacting pairs occur on the outside of the barrel, as do 55 of the
62 Aromatic-Aromatic pairs that form non-H-bonded interactions.  The
preference of Gly-Gly pairs to face the inside of the barrel is also
significant for non-H-bonded interacting pairs (17 out of 19 occur on
the inside).  However, no clear preference is seen for Gly-Gly strong
H-bonds: Only 8 of 16 pairs occur on the inside.  For strong H-bonded
pairs between Gly and aromatic residues, 48 out of 63 pairs are
internal, whereas there is no preference for non-H-bonded pairs (14
out of 25 are internal).  This may reflect a preference for ``aromatic
rescue'' of Gly, discussed below.  The orientation for weak H-bonded
pairs is not relevant, since one residue always faces the inside of
the barrel and the other always faces the outside, as shown in
Figure~\ref{Fig:interaction}.

\subsubsection{Comparison to soluble barrel and barrel-like $\beta$-sheets}
In order to determine whether the pairwise propensities calculated for
TM $\beta$-barrels were due to the effects of the transmembrane
environment or due to the intrinsic structural effects of
$\beta$-barrels, we calculated pairwise propensities for a set of
soluble $\beta$-barrels and barrel-like $\beta$-sheets that were
structurally similar to the TM $\beta$-barrel dataset as determined by
the Combinatorial Extension structural alignment method
\cite{Shindyalov98_PE}.  The soluble dataset contains 28 proteins.
All of the proteins contain anti-parallel $\beta$-sheets with a
right-hand twist, similar to TM $\beta$-barrels.  Some open
$\beta$-barrels and other $\beta$-sheets were included to ensure that
the size of the dataset would be similar to that of the TM
$\beta$-barrels.  The final dataset has approximately the same number
of strand pairs, though about 40\% fewer residue pairs.  Significant
pair propensities for the soluble $\beta$-sheet dataset for each
interaction type are listed in Supplementary Material.

Only two motifs appear in both TM and soluble proteins: G-V non-bonded
pairs (propensity 1.60 in TM $\beta$-barrels, 1.78 in soluble
$\beta$-sheets) and G-P weak H-bonds (1.78 TM, 3.75 soluble).  In
soluble $\beta$-sheets, polar-polar and hydrophobic-hydrophobic pairs
have high propensities for strong H-bond and non-H-bonded pairings,
while polar-hydrophobic pairs have low propensities.  The results for
weak H-bonds show an opposing trend.  The strongest motifs in TM
$\beta$-barrels, W-Y non-H-bonded pairs and G-Y strong H-bonds, are
not statistically significant in soluble sheets.  W-Y non-H-bonded
pairs have propensity 1.06 and G-Y strong H-bonds have propensity 1.27
with $p$-value 0.55.

Pairwise interhelical interactions in TM $\alpha$-helical membrane
proteins have also been studied using odds ratios
\cite{Adamian01_JMB}.  The only favorable interaction motif found in
both TM $\alpha$-helices and TM $\beta$-barrels is G-F, a strong
H-bond motif (odds ratio 1.80) involved in aromatic rescue.
Otherwise, there are very few similarities between the two protein
families.

\subsection{Prediction of strand register}
The single and pairwise propensities can be useful for structure
prediction.  We have developed an algorithm to predict strand register
in $\beta$-barrel membrane proteins knowing only amino acid sequences
and the approximate positions of the first residue in each strand.
The latter can be obtained by several strand predictors, though they
cannot themselves predict strand register.  We use the hidden Markov
model predictor of Bigelow {\it et al.}  \cite{Bigelow04_NAR}.  We
also test our method when the actual strand starts are given
beforehand.

Our predictor consists of two steps.  In step 1, we use the
single-body regional propensities to refine the approximate strand
starts given by the HMM predictor into ``true'' strand starts.  It is
only necessary for the strand predictor to predict strand starts
within 5 residues of the exact strand start.  In step 2, we use the
two-body interstrand propensities ({\sc Tmsip}) to predict final
strand register for each strand pair.  Details of the algorithm are
discussed in Methods.

\renewcommand{\baselinestretch}{1.0}
\begin{table}[t!] 
\begin{center}
      \vspace{4mm} 
\begin{tabular}{lcrr}
      \hline \hline 
Method & & Strands & Accuracy \\
\cline{1-1} \cline{3-4}

Full prediction & & 112 & 44\% \\
Step 1 only     & & 100 & 39\% \\
Step 2 only     & &  92 & 36\% \\
No prediction   & &  61 & 24\% \\
Random          & &  17 &  7\% \\
      \hline \hline 

\end{tabular}
\end{center}
      \caption{\small Prediction of strand register in $\beta$-barrel
membrane proteins.  We report the success of our tests as the number
of strand pairs whose register is correctly predicted (``Strands'')
and the proportion out of the 256 strand pairs tested (``Accuracy'').
Listed are the accuracies for using both steps (single and pairwise
propensities) as described in Methods (``Full prediction''), using
step 1 or 2 individually, skipping both steps (``No prediction''), or
using randomly generated strand registers (``Random'').  }
\label{tab:prediction}
\end{table}

The results of our predictor are reported in
Table~\ref{tab:prediction}, where the proportion of strand pairs (out
of 256) whose register is correctly predicted is listed.  When the
strand starts predicted by the method of Bigelow {\it et al.}\ are
given as input, the accuracy is 44\%.  When the actual strand starts
from the PDB structures are given as input instead, the accuracy is
46\%.  This shows that only approximate strand starts are necessary
for strand registration prediction.

We compare our accuracy to a random control, in which we randomly
select a window from the 11 available positions for each strand and
then randomly select a strand shear from the 11 possible values for
each strand pair (as described in Methods).  We ran this random
selection 10 times on the dataset, and only 17.4 strand pairs were
predicted on average (standard deviation 3.61).  This translates to a
random accuracy of 7\%.  Our accuracy of 44\% is thus considerably
better than random.

We also compare our results to a similar prediction of strand register
in soluble $\beta$-sheets by Steward and Thornton
\cite{Steward02_Prot}.  This prediction uses the sum of single and
pairwise scores to align one $\beta$-strand against a fixed strand
that it is known to pair with, and achieves an accuracy of 31\% in
antiparallel $\beta$-sheets.  These conditions are similar to ours
when the actual strand starts are given as inputs and we skip step 1.
We achieve an accuracy of 68\% in TM $\beta$-barrels under these
conditions.

\section{Discussion}

We have described residue preferences for different regions and a
number of spatial patterns for $\beta$-barrel membrane proteins.  Many
of the novel patterns discovered in this study provide useful
information about the assembly and folding process of $\beta$-barrel
membrane proteins.

\paragraph*{``Positive-outside'' rule.}
Basic residues have an asymmetric distribution between the two cap
regions of $\beta$-barrel membrane proteins.  Arg and Lys have
propensities of 0.61 and 0.86 in the periplasmic cap region,
respectively, but 1.54 and 1.54 in the extracellular cap region, and
thus have over twice the preference to be found in the extracellular
cap region than in the periplasmic cap region.  This is notable for
several reasons: first, one would intuitively expect charged residues
to have high propensity in both cap regions, since they are surrounded
by an aqueous environment; second, acidic residues have high and
relatively similar propensities in both cap regions (Asp has a
propensity of 1.83 for the periplasmic cap and 1.53 for the
extracellular cap, and Glu has 1.20 and 1.32 for these regions,
respectively); third, despite their poor propensity for the
periplasmic cap, basic residues have very high propensities for the
internal face of the periplasmic headgroup region (Arg 1.62, Lys
2.37), and often ``snorkel'' away from the center of the lipid bilayer
and extend into the periplasmic space.

A similar asymmetric distribution of basic residues in cap regions has
been found in $\alpha$-helical membrane proteins, yet the distribution
is reversed: Basic residues have a high propensity for the inner,
cytoplasmic cap, and a low propensity for the outer cap (facing the
extracellular space in eukaryotes and gram-positive bacteria or facing
the periplasm in gram-negative bacteria) \cite{vonHeijne89_Nat}.  This
property was named the ``positive-inside'' rule.  By analogy,
$\beta$-barrel membrane proteins thus show a ``positive-outside''
rule.

A possible explanation for this result is that the asymmetric
distribution of basic residues is related to the asymmetric lipid
composition of the outer membrane.  The inner leaflet of the outer
membrane, which faces the periplasm, is composed primarily of
phosphatidylethanolamine (PE), while the outer leaflet, which faces
the extracellular environment, has a high concentration of
lipopolysaccharide (LPS), which bears several negatively charged
groups.  The asymmetric distribution of basic residues within the
protein may lead to a stable positioning of the protein within this
asymmetric lipid bilayer.  It has been suggested that $\beta$-barrel
membrane proteins may interact with LPS in the periplasm before
co-inserting into the outer membrane \cite{deCock96_EMBO}.  If this is
true, the high propensity of basic residues on the extracellular side
of the protein would attract the highly negatively-charged LPS so that
the protein will insert in the correct direction.  In a recent study
in which the structure of the TM $\beta$-barrel of FhuA is
co-crystallized with one LPS molecule, LPS binds strongly to basic
residues on FhuA in the extracellular cap region, forming hydrogen
bonds or ionic interactions \cite{Ferguson98_Sci,Nikaido03_MMBR}.
Another study suggests that the frequency of positively charged
residues in the headgroup region is correlated to {\it in vitro}
affinity of $\beta$-barrel membrane proteins for different
phospholipids. \cite{Ramakrishnan04_BC}.  Thus, the affinity of the
two sides of the barrel for the two outer membrane leaflets may be
affected by the amino acid composition of the barrel and the lipid
composition of the leaflets.

\paragraph*{Aromatic rescue of glycine in $\beta$-barrel membrane proteins.}
Tyr-Gly is the most favorable interstrand motif for backbone H-bond
interactions (propensity 1.56, $p$-value 8$\times 10^{-4}$).  Of the
39 G-Y strong H-bonded pairs in the dataset, 32 are on the inside of
the barrel, where Tyr is normally disfavored.  Tyr adopts an unusual
rotamer (60,90) in this location: 60\% of internal Tyr side-chains
adopt this rotamer, while only 8\% of external Tyr side-chains and 6\%
of Tyr side-chains in soluble $\beta$-strands do
\cite{Chamberlain04_BJ}.  There is a single explanation for this
rotamer preference and the unusual preference of Tyr for the inside of
the barrel when it is part of a G-Y interaction.  As shown in
Figure~\ref{Fig:inter}b, the side-chain of Tyr has extensive contacts
with and covers the Gly residue to which it is backbone H-bonded.  Of
the 32 internal G-Y strong H-bonded pairs, 30 show this behavior, but
only 1 of 7 external G-Y pairs do.

Merkel and Regan discovered the same behavior in soluble
$\beta$-sheets, and named it ``aromatic rescue'' of glycine
\cite{Merkel98_FD}.  They found that aromatic rescue mitigates the
instability Gly causes in $\beta$-sheets.  They proposed that the
aromatic side-chain prevents the exposure of the backbone around Gly
to solvent while at the same time minimizing the surface area of the
aromatic ring exposed to solvent.  This effect also accounts for the
fact that this pattern occurs preferentially on the inside of the
barrel in TM regions, where it is exposed to solvent or a polar
environment.  Phe behaves similarly to Tyr in this region.  G-F is a
significant strong H-bonded motif (propensity 1.80, $p$-value 5$\times
10^{-3}$) and displays aromatic rescue on the inside of the barrel.
Likewise, the (60,90) rotamer is observed in 83\% of internal Phe
side-chains, but only in 19\% of external Phe side-chains and 5\% of
Phe side-chains in soluble $\beta$-strands \cite{Chamberlain04_BJ}.

Hutchinson {\it et al.} studied the relationship between H-bonded and
non-H-bonded pairs in soluble $\beta$-sheets, and found that G-F
H-bonds were 2.63 times more likely than G-F non-H-bonded pairs
\cite{Hutchinson98_PS}.  This is similar to the ratio for TM
$\beta$-barrels (2.37).  However, the ratio of G-Y for TM
$\beta$-barrels (2.93) is considerably higher than that for soluble
$\beta$-sheets (1.39).  Tyr may have a much larger role in aromatic
rescue in TM $\beta$-barrel membrane proteins than soluble
$\beta$-sheets because of the higher polarity of Tyr side-chains
compared to Phe.  Aromatic rescue may be even more important for the
internal surface of TM $\beta$-barrels than for soluble
$\beta$-sheets, both because of the need to have polar residues such
as Tyr on the internal surface and the need to have Gly in order to
relieve the steric pressure caused by the curvature of the barrel.

\paragraph*{Structure prediction.}
We have incorporated our propensity values into a preliminary study
predicting strand pair register in TM $\beta$-barrels.  We find that
our predictor can achieve significantly higher accuracy than random.
It can achieve nearly the same accuracy by inputting strand starts
approximated by the hidden Markov model predictor of Bigelow {\it et
al.}\ \cite{Bigelow04_NAR} as when the true strand starts are known.
Thus, only the amino acid sequence is necessary for strand pairing
prediction if the method of Bigelow {\it et al.}\ is used first.  Our
accuracy is considerably better than random (44\% {\it vs.}\ 7\%), and
is also better than just relying on the initial HMM prediction (24\%).
The predictor discussed in this study represents the first step
towards a full structure prediction of TM $\beta$-barrels.

We also assessed the effects of using only knowledge of strand starts
(step 1) or only strand pairing (step 2).  We can skip step 1 by not
refining the strand starts given as input (``Step 2 only'' in
Table~\ref{tab:prediction}).  We find that the accuracy of the
predictor decreases to 36\% using start positions from the HMM
predictor.  This shows that the single-body regional preferences
improve accuracy by 8\%.  We can skip step 2 (``Step 1 only'' in
Table~\ref{tab:prediction}) by not using our pairwise interstrand
propensities, and simply assigning the most frequently observed strand
register as correct.  In this case, the accuracy decreases to 39\%.
Clearly, both steps 1 and 2 are necessary to achieve maximal accuracy.
When both steps 1 and 2 are skipped (``No prediction'' in
Table~\ref{tab:prediction}), the accuracy drops to 24\%.

\paragraph*{Determining factors of folding and assembly.}
There are two main factors that determine the spatial patterns
described in this study.  As with $\alpha$-helical membrane proteins,
the folding and assembly of $\beta$-barrel membrane proteins in the
transmembrane region are governed by the thermodynamics of lipids and
proteins and their interactions.  On the other hand, the sorting and
targeting process required for integrating membrane proteins into
lipid bilayers in the correct direction after biosynthesis is an
important process of the cell, and involves complex biomolecular
machinery \cite{Tamm04_BBA}.  For helical membrane proteins, the
nature of such machinery is being uncovered by the study of
translocons \cite{White04_COSB}.  For $\beta$-barrel proteins, recent
work on chaperone-assisted folding hint at a very complex machinery as
well \cite{Tamm04_BBA}.

Distinguishing spatial patterns due to fundamental thermodynamics from
those due to the unique requirements of biological localization is a
challenging task.  The identification and understanding of the origins
of these motifs will help to elucidate the folding mechanisms of
$\beta$-barrel proteins.  An attractive hypothesis is that these two
types of events can be discerned from further analysis of
propensities, motifs, and antimotifs.

\paragraph*{Experimentally testable hypotheses.}
The identification of favorable pairing residues may be useful for
suggesting experimental studies to elucidate determinants important
for {\it in vivo\/} folding and to dissect the determinants of
thermodynamic stability for $\beta$-barrel membrane proteins.  For
example, Arg and Lys can be introduced in the periplasmic cap region,
and if such mutants can fold with similar stability in {\it in
vitro\/} experiments, this would suggest that the ``positive-outside''
rule may be important mostly for {\it in vivo\/} folding.  In
addition, measurement of the thermodynamic stabilities of
$\beta$-barrel proteins in reconstituted lipid bilayers of different
composition can help to clarify whether the origin of the
``positive-outside'' rule is due to the asymmetric composition of
phospholipid bilayers.

Gly-Tyr interstrand pairs involved in aromatic rescue may serve as
anchoring sites of $\beta$-barrel membrane protein folding.  To study
details of the folding mechanism, one could remove and add anchoring
Gly-Tyr residue pairs at strategic positions simultaneously in order
to introduce maximal changes in the effective contact orders, which
would significantly alter the zipping process of folding
\cite{Weikl03_JMB}.  Folding rate studies of these mutated proteins
would be valuable for elucidating the folding mechanism of
$\beta$-barrel proteins in general and the role of the high-propensity
Gly-Tyr interaction in particular.  The motifs,
antimotifs, and spatial high propensity pairing described in this
study can be used profitably for studying the folding and assembly
mechanism of $\beta$-barrel membrane proteins.

\paragraph*{Importance of statistical analysis for small datasets.}
Previous studies of residue pairing in soluble $\beta$-sheets either
lack a statistical model and hence provide no statistical
significance in their results, or use simple models such as the
$\chi^2$ test for $p$-value calculation, which is inappropriate for
studying the pairing of very short strands
\cite{Steward02_Prot,Wouters95_Prot,Hutchinson98_PS}.

Because no existing textbook or published statistical models are
applicable for analyzing short strand pairing, the key component of
our method is the development of methods of exact calculation of
$p$-values for the null model of exhaustive permutation of strand
sequences.  With the development of a higher-order generalized
hypergeometric model based on the combinatorics of short sequences, we
are able to identify very significant motifs and antimotifs from very
limited data.  The development of rigorous statistical methods is an
important technical development of work reported in this paper.

\section{Model and Methods}
We sketch briefly our models and computational methods below.  More
 details can be found in {\it Supplementary Material}).
\paragraph*{Database.}
The dataset used to derive the statistical models comprises 19
$\beta$-barrel membrane proteins found in the Protein Data Bank
(Table~\ref{tab:dataset}), totaling 262 $\beta$-strands.  All proteins
share no more than 26\% pairwise sequence identity.  All structures have a
resolution of 2.6 \AA\ or better.

\renewcommand{\baselinestretch}{1.0}
\begin{table}[t!] 
\begin{center}
      \vspace{4mm} 
\begin{tabular}{lllrl}
      \hline \hline 
\multicolumn{1}{c}{Protein} & \multicolumn{1}{c}{Organism} & \multicolumn{1}{c}{Architecture} &
\multicolumn{1}{c}{Strands} & \multicolumn{1}{c}{PDB ID} \\ \hline

OmpA & {\it E. coli} & monomer & 8 & \tt 1BXW \cite{Pautsch98_NSB} \\
OmpX & {\it E. coli} & monomer & 8 & \tt 1QJ8 \cite{Vogt99_SFD} \\
NspA & {\it N. meningitidis} & monomer & 8 & \tt 1P4T \cite{Vandeputte03_JBC} \\
OpcA & {\it N. meningitidis} & monomer & 10 & \tt 1K24 \cite{Prince02_PNAS} \\
OmpT & {\it E. coli} & monomer & 10 & \tt 1I78 \cite{Vandeputte01_EMBO} \\
OMPLA & {\it E. coli} & dimer & 12 & \tt 1QD6 \cite{Snijder99_Nat} \\
NalP & {\it N. meningitidis} & monomer & 12 & \tt 1UYN \cite{Oomen04_EMBO} \\
Porin & {\it R. capsulatus} & trimer & 16 & \tt 2POR \cite{Weiss92_JMB} \\
Porin & {\it R. blastica} & trimer & 16 & \tt 1PRN \cite{Kreusch94_JMB} \\
OmpF & {\it E. coli} & trimer & 16 & \tt 2OMF \cite{Cowan95_Struct} \\
Omp32 & {\it C. acidovorans} & trimer & 16 & \tt 1E54 \cite{Zeth00_SFD} \\
LamB & {\it S. typhimurium} & trimer & 18 & \tt 2MPR \cite{Meyer97_JMB} \\
ScrY & {\it S. typhimurium} & trimer & 18 & \tt 1A0S \cite{Forst98_NSB} \\
FepA & {\it E. coli} & monomer & 22 & \tt 1FEP \cite{Buchanan99_NSB} \\
FhuA & {\it E. coli} & monomer & 22 & \tt 2FCP \cite{Ferguson98_Sci} \\
FecA & {\it E. coli} & monomer & 22 & \tt 1KMO \cite{Ferguson02_Sci} \\
BtuB & {\it E. coli} & monomer & 22 & \tt 1NQE \cite{Chimento03_NSB} \\
TolC & {\it E. coli} & trimer & 4 & \tt 1EK9 \cite{Koronakis00_Nat} \\
$\alpha$-Hemolysin & {\it S. aureus} & heptamer & 2 & \tt 7AHL \cite{Song96_Sci} \\
      \hline \hline 

\end{tabular}
\end{center}
      \caption{\small Dataset of 19 $\beta$-barrel membrane proteins
used for this study.  All proteins share no more than 26\% pairwise
sequence identity.  Crystal structures have a resolution of 2.6 \AA\
or less.  Three identical chains of TolC and seven of
$\alpha$-hemolysin form a single barrel; all other proteins listed
form whole barrels with a single peptide chain.}  \label{tab:dataset}
\end{table}

The dataset of soluble barrel and barrel-like $\beta$-sheets was
compiled by searching structural homologs using the Combinatorial
Extension method \cite{Shindyalov98_PE} with each of the 19 proteins
in the TM dataset.  Because the resulting dataset was small, CE was
used again on the two closed barrels obtained from the first attempt,
streptavidin (pdb {\tt 1stp}) and retinol binding protein (pdb {\tt
1brp}).  As with the TM dataset, all proteins share no more than 26\%
pairwise sequence identity.  All structures have a resolution of 3.3
\AA\ or better.  The dataset comprises 28 soluble $\beta$-sheets: {\tt
1avg}, {\tt 1ayr}, {\tt 1bbp}, {\tt 1bj7}, {\tt 1bpo}, {\tt 1brp},
{\tt 1dfv}, {\tt 1dmm}, {\tt 1d2u}, {\tt 1eg9}, {\tt 1ei5}, {\tt
1epa}, {\tt 1em2}, {\tt 1ewf}, {\tt 1fsk}, {\tt 1fx3}, {\tt 1h91},
{\tt 1jkg}, {\tt 1lkf}, {\tt 1m6p}, {\tt 1qfv}, {\tt 1std}, {\tt
1stp}, {\tt 1t27}, {\tt 1una}, {\tt 2a2u}, {\tt 3blg}, {\tt 4bcl}.

\paragraph*{Single-body propensities.}
We define the single-body propensity $P_r(X)$ of residue type $X$ in
region $r$ as the odds ratio comparing the frequency of a residue type
in one region to its expected frequency when all eight regions are
combined:
\[
P_r(X) = \frac{f(X|r)}{\mathbb{E}[f'(X|r)]},
\]
where $f(X|r)$ is the observed frequency (number count) of residue
type $X$ in region $r$, and $\mathbb{E}[f'(X|r)]$ is the
expected frequency of residue type $X$ in region $r$.

In order to calculate $\mathbb{E}[f'(X|r)]$, we use a random null
model of exhaustive permutation of all residues in all eight
regions. Here each permutation occurs with equal probability.  
The probability $\mathbb{P}_{X|r}(i)$ of $i=f'(X|r)$
residues of type $X$ being assigned to region $r$ follows a
hypergeometric distribution:
$
\mathbb{P}_{X|r}(i) 
= h(i | n, n_r, n_x) = \binom{n_x}{i}\binom{n-n_x}{n_r-i}/\binom{n}{n_r},
$
where $n$ is the number of residues of all types in all eight regions
in the entire dataset, $n_r$ is the number of residues of all types
in region $r$, and $n_x$ is the number of residues of type $X$ in all
regions.  
 The expected frequency of residue type $X$ occurring in region
$r$ is the mean of the hypergeometric distribution $h(i | n, n_r, n_x)$:
\begin{equation}
\mathbb{E}[f'(X|r)] = \sum_{i=0}^{n_x}i \cdot \mathbb{P}_{X|r}(i)  = \frac{n_r \cdot n_x}{n}.
\label{eqn:hyper}
\end{equation}
Therefore, the propensity $P_r(X)$ is:
\[
P_r(X) = \frac{f(X|r)}{\mathbb{E}[f'(X|r)]} = \frac{f(X|r) / n_r}{n_x / n}.
\]

Because the frequency of occurrences of residues of type $X$ in region
$r$ in the null model follows a hypergeometric distribution, we can
calculate an exact $p$-value for an observed $f(X|r)$ to assess its
statistical significance.  We calculate a two-tailed $p$-value based
on the null hypothesis that $P_r(X)=1.0$.

\paragraph*{Determination of pairwise contacts.}
The three types of pairwise contacts (strong H-bond, side-chain, and
weak H-bond) were assigned for interacting residues based on contact
types defined by the DSSP program (Definition of Secondary Structure
of Proteins \cite{Kabsch83_Biopoly}) based on the atomic coordinates
from PDB files.  For $\beta$-sheets, DSSP defines {\it bridge partners}
as residues across from each other on adjacent $\beta$-strands, and
also determines whether bridge partners interact via backbone N-O
H-bonds.  Two TM residues on adjacent $\beta$-strands contribute to a
backbone {\it strong H-bond} interaction if they are listed as bridge
partners that contribute to backbone H-bonds.  They contribute to a
{\it non-H-bonded interaction} if they are listed as bridge partners
but do not contribute to backbone H-bonds.  They contribute to a {\it
weak H-bond} if the residue with the smaller residue number ({\it
i.e.} closer to the N-terminus of the protein) is a bridge partner to
the residue immediately following the other residue of the pair.
This is because a weak C$_\alpha$-O H-bond extends by one residue in
the N-C direction, as seen in Figure~\ref{Fig:interaction}.  Residues
$j$ and $y$ contribute to a weak H-bond in
Figure~\ref{Fig:interaction}, for example, because the residue with
the lower number ($y$, since it is closer to the N-terminus), is a
bridge partner to $i$, the residue immediately following $j$.  For a
strand pair between the first and last strands of a protein, the
larger and smaller numbered residues must be switched, to account for
the fully circular nature of $\beta$-barrels.  Residues outside the TM
regions (core and headgroup) were not considered in our calculations
of pairwise contacts.

\paragraph*{Interstrand two-body spatial contact propensities.}
The interstrand pairwise propensity ({\sc Tmsip}) $P(X,Y)$ of residue
types $X$ and $Y$ for each of the three types of pairwise contacts is
given by:
\[
P(X,Y) = \frac{f(X,Y)}{\mathbb{E}[f'(X,Y)]},
\]
where $f(X,Y)$ is the observed frequency of $X$-$Y$ contacts of a
specific type in the TM regions (core and headgroup), and
$\mathbb{E}[f'(X,Y)]$ is the expected frequency of $X$-$Y$ contacts in
a null model.  In Tables~\ref{tab:interSig} and \ref{tab:interWeight},
$f(X,Y)$ is listed under ``Obs.'', and
$\mathbb{E}[f'(X,Y)]$ is listed under ``Exp.''

In order to calculate $\mathbb{E}[f'(X,Y)]$, we choose a null model in
which residues within each of the two adjacent strands in a strand
pair are permuted exhaustively and independently, and each permutation
occurs with equal probability.

{\it Contacts between residues of the same type.\/} When $X$ is the
same as $Y$, the probability $\mathbb{P}_{X,X}(i) $
of $i=f'(X,X)$ number of $X$-$X$ contacts
in a strand pair follows a hypergeometric distribution.
$\mathbb{E}_{\rm all}[f'(X,X)]$ is then the sum of the expected values
of $f'(X,X)$ for the set ${\cal SP}$ of all strand pairs in the
dataset:
\[
\mathbb{E}_{\rm all}[f'(X,X)] 
= \sum_{sp \in {\cal SP}} \mathbb{E}[f'_{sp}(X,X)]
= \sum_{sp \in {\cal SP}}
\frac{x_1(sp)\cdot x_2(sp)}{l(sp)},
\]
where $x_1(sp)$ and $x_2(sp)$ are the numbers of residues of type $X$
in the first and second strand of strand pair $sp \in {\cal SP}$,
respectively, and $l(sp)$ is the length of strand pair $sp$.  The
right-hand side is determined by using the expectation of the
hypergeometric distribution, analogous to Equation~(\ref{eqn:hyper}).
For statistical significance, two-tailed $p$-values can be calculated
using $\mathbb{P}_{X,X}$.

{\it Contacts between residues of different types.\/} If the two
contacting residues are not of the same type, {\it i.e.}\ $X \neq Y$,
then the number of $X$-$Y$ contacts in the random model for one strand
pair is the sum of two dependent hypergeometric variables, one
variable for type $X$ residues in the first strand and type $Y$ in the
second strand, and another variable for type $Y$ residues in the first
strand and type $X$ in the second strand.  The expected frequency of
$X$-$Y$ contacts $\mathbb{E}[f'(X,Y)]$ is the sum of the two expected
values over all strand pairs $sp \in {\cal SP}$:
\[
\mathbb{E}[f'(X,Y)] 
= \sum_{sp \in {\cal SP}}\{
\mathbb{E}[f'_{sp}(X,Y)] +
\mathbb{E}[f'_{sp}(Y,X)]\}
= \sum_{sp \in {\cal SP}}\{
\frac{x_1(sp)\cdot y_2(sp)}{l(sp)} + \frac{y_1(sp)\cdot x_2(sp)}{l(sp)}
\},
\]
where $x_1(sp)$ and $x_2(sp)$ are the numbers of residues of type $X$
in the first and second strand, $y_1(sp)$ and $y_2(sp)$ are the
numbers of residues of type $Y$ in the first and second strand, and
$l(sp)$ is the length of strand pair $sp$.  The right-hand side is
determined by using the expectation of the hypergeometric
distribution, analogous to Equation~(\ref{eqn:hyper}).  A more complex
hypergeometric formula for the null model is used for exact
calculation of $p$-values (see details in {\it Supplementary
Material}).

\paragraph*{Strand register prediction.}
Our algorithm consists of two steps, the prediction of exact strand
starts and the prediction of strand register.  In both steps, a probabilistic
model is used which involves the summation of log-propensity values
calculated in this study.

For step 1, we use the single-body regional propensities.  We begin
with an initial postion of strand start, {\it e.g.}, as predicted by
the hidden Markov model of Bigelow {\it et al.}  \cite{Bigelow04_NAR},
and sum the log-propensities of the surrounding amino acids for 11
windows: the window starting at the predicted strand start, and all
windows within 5 residues from the original prediction.  The window
with the sum that reflects the highest probability is taken as the
``true'' strand start used in step 2.

For step 2, we use the pairwise two-body propensities.  We begin with
two adjacent strands, using the strand starts predicted in step 1.  We
then shift one strand against another, and sum the log-propensities of
the surrounding residue pairs for 11 windows (called ``strand
shears''): the window in which the two strand starts are bonded
together, and all windows within 6 residues up and 4 residues down for
the original position.  This is because a strand shear of +1 is the
most commonly occurring strand shear in the dataset.  The window with
the sum that reflects the highest probability is taken as the correct
strand pairing.

We exclude from our prediction analysis two proteins in our dataset (6
strand pairs) for which the HMM strand start predictor failed: TolC
and $\alpha$-HL.  This leaves our dataset with 17 proteins comprising
256 strands.  We test our method in ``leave-one-out'' fashion, using
16 proteins to derive the log-propensity scores use to predict strand
pairing in the 17th protein.  The results are listed in
Table~\ref{tab:prediction}.

\section{Acknowledgment}
We thank Dr.\ Bosco Ho for insightful comments.  We thank Xiang Li,
Drs.\ William Wimley and Jinfeng Zhang for helpful discussions.  We
thank Sarah Cheng for programming assistance.  This work is supported
by grants from the National Science Foundation (CAREER DBI0133856 and
DBI0078270), National Institute of Health (GM68958), and Office of
Naval Research (N000140310329).

\section{Supplementary Material}

This is a self-containing expaned section on methods.

\paragraph*{Database.}
The dataset used to derive the statistical models comprises 19
$\beta$-barrel membrane proteins found in the Protein Data Bank
(Table 5), totaling 262 $\beta$-strands.  All proteins
share no more than 26\% pairwise sequence identity.  All structures have a
resolution of 2.6 \AA\ or better.

The dataset of soluble barrel and barrel-like $\beta$-sheets was
compiled by searching structural homologs using the Combinatorial
Extension method \cite{Shindyalov98_PE} with each of the 19 proteins
in the TM dataset.  Because the resulting dataset was small, CE was
used again on the two closed barrels obtained from the first attempt,
streptavidin (pdb {\tt 1stp}) and retinol binding protein (pdb {\tt
1brp}).  As with the TM dataset, all proteins share no more than 26\%
pairwise sequence identity.  All structures have a resolution of 3.3
\AA\ or better.  The dataset comprises 28 soluble $\beta$-sheets: {\tt
1avg}, {\tt 1ayr}, {\tt 1bbp}, {\tt 1bj7}, {\tt 1bpo}, {\tt 1brp},
{\tt 1dfv}, {\tt 1dmm}, {\tt 1d2u}, {\tt 1eg9}, {\tt 1ei5}, {\tt
1epa}, {\tt 1em2}, {\tt 1ewf}, {\tt 1fsk}, {\tt 1fx3}, {\tt 1h91},
{\tt 1jkg}, {\tt 1lkf}, {\tt 1m6p}, {\tt 1qfv}, {\tt 1std}, {\tt
1stp}, {\tt 1t27}, {\tt 1una}, {\tt 2a2u}, {\tt 3blg}, {\tt 4bcl}.

\paragraph*{Spatial Regions and Strand Model.}
In order to classify residues in $\beta$-barrel membrane proteins by
specific spatial regions (core, headgroup, and polar caps), the
coordinates in the protein's PDB file were translated and rotated so
that the $xy$-plane was perpendicular to the vertical axis of the
barrel and equidistant to the observed aromatic girdles presumed to be
at the membrane interfaces.  An example of such a transformation is
illustrated in Figure 1.  Each residue in the protein
was assigned a region based on the $z$-coordinate of its associated
$\alpha$-carbon.

In addition, TM residues (those in the core or headgroup regions) were
assigned as {\it internal} ({\it i.e.}\ their side-chains face into
the center of the barrel) or {\it external} ({\it i.e.}\ their
side-chains face away from the center of the barrel), depending on the
angle between the vector from the barrel central axis to the
$\alpha$-carbon and the vector from the $\alpha$-carbon to the
$\beta$-carbon of the residue.  If the angle is less than 90$^\circ$,
it is classified as internal; if the angle is greater, it is
classified as external.  If the residue is glycine, its orientation
is extrapolated as the opposite of the residue previous to it on the
$\beta$-strand.

In total, each barrel is divided into 8 distinct regions: periplasmic
cap with $z \in (-$20.5\AA, $-$13.5\AA), periplasmic headgroup (internal and
external) with $z \in (-$13.5\AA, $-$6.5\AA), core (internal and external) with
$z \in (-$6.5\AA, 6.5\AA), extracellular headgroup (internal and external)
with $z \in ($6.5\AA, 13.5\AA), and extracellular cap with $z \in
($13.5\AA, 20.5\AA).  Residues that are not assigned to be in $\beta$-strands
are automatically excluded from the TM (headgroup and core) regions.
Unless noted, any residue in the protein not assigned to any of these
regions was excluded from the calculation.

\paragraph*{Single-body propensities.}
We define the single-body propensity $P_r(X)$ of residue type $X$ in
region $r$ as the odds ratio comparing the frequency of a residue type
in one region to its expected frequency when all eight regions are
combined:
\[
P_r(X) = \frac{f(X|r)}{\mathbb{E}[f'(X|r)]},
\]
where $f(X|r)$ is the observed frequency (number count) of residue
type $X$ in region $r$, and $\mathbb{E}[f'(X|r)]$ is the
expected frequency of residue type $X$ in region $r$.

In order to calculate $\mathbb{E}[f'(X|r)]$, we need a random null
model.  Here we chose as our model exhaustive permutation of all
residues in all eight regions, such that each permutation occurs with
equal probability.  That is, the residues in all eight regions of all
proteins in the dataset are permuted exhaustively, without
replacement, and assigned regions based on their new positions.  For
each permutation, $f'(X|r)$ records the number of occurrences of residue
type $X$ in region $r$.  Under this model, the probability
$\mathbb{P}_{X|r}(i)$ of $i=f'(X|r)$ residues of type $X$ being
assigned to region $r$ follows a hypergeometric distribution:
\[
\mathbb{P}_{X|r}(i) 
= h(i | n, n_r, n_x) = \frac{\binom{n_x}{i}\binom{n-n_x}{n_r-i}}{\binom{n}{n_r}},
\]
where $n$ is the number of residues of all types in all eight regions
in the entire dataset, $n_r$ is the number of residues of all types
in region $r$, and $n_x$ is the number of residues of type $X$ in all
regions.  Analogously, we can think of $\mathbb{P}_{X|r}(i)$ as the probability of
selecting without replacement $n_r$ residues out of a total of $n$
residues contained in an urn, such that $i$ of the residues are of type
$X$.  The expected frequency of residue type $X$ occurring in region
$r$ is the mean of the hypergeometric distribution $h(i | n, n_r, n_x)$:
\begin{equation}
\mathbb{E}[f'(X|r)] = \sum_{i=0}^{n_x}i \cdot \mathbb{P}_{X|r}(i)  = \frac{n_r \cdot n_x}{n}.
\label{eqn:hyper2}
\end{equation}
Therefore, the propensity $P_r(X)$ is:
\[
P_r(X) = \frac{f(X|r)}{\mathbb{E}[f'(X|r)]} = \frac{\frac{f(X|r)}{n_r}}{\frac{n_x}{n}}.
\]
That is, $P_r(X)$ is the ratio of the proportion of residue type $X$
in region $r$ to the proportion of residue type $X$ in all eight
regions combined.

Because the frequency of occurrences of residues of type $X$ in region
$r$ in the null model follows a hypergeometric distribution, we can
calculate an exact $p$-value for an observed $f(X|r)$ to assess its
statistical significance.  We calculate a two-tailed $p$-value based
on the null hypothesis that $P_r(X)=1.0$.  If the observed
$P_r(X)<1.0$, then:
\begin{equation}
p = 2 \cdot \sum_{i=0}^{f(X|r)}\mathbb{P}_{X|r}(i).
\label{eqn:lowpvalue}
\end{equation}
If the observed $P_r(X)>1.0$, then:
\begin{equation}
p = 2 \cdot \sum_{i=f(X|r)}^{x_n}\mathbb{P}_{X|r}(i).
\label{eqn:highpvalue}
\end{equation}
That is, the $p$-value is the probability that the propensity
calculated from the dataset would deviate as much or more from the
observed propensity, higher or lower, assuming that the actual
propensity is 1.0.

\paragraph*{Determination of pairwise contacts.}
The three types of pairwise contacts (strong H-bond, side-chain, and
weak H-bond) were assigned for interacting residues based on contact
types defined by the DSSP program (Definition of Secondary Structure
of Proteins \cite{Kabsch83_Biopoly}) based on the atomic coordinates
from PDB files.  For $\beta$-sheets, DSSP defines {\it bridge partners}
as residues across from each other on adjacent $\beta$-strands, and
also determines whether bridge partners interact via backbone N-O
H-bonds.  Two TM residues on adjacent $\beta$-strands contribute to a
backbone {\it strong H-bond} interaction if they are listed as bridge
partners that contribute to backbone H-bonds.  They contribute to a
{\it non-H-bonded interaction} if they are listed as bridge partners
but do not contribute to backbone H-bonds.  They contribute to a {\it
weak H-bond} if the residue with the smaller residue number ({\it
i.e.} closer to the N-terminus of the protein) is a bridge partner to
the residue immediately following the other residue of the pair.
This is because a weak C$_\alpha$-O H-bond extends by one residue in
the N-C direction, as seen in Figure 2.  Residues
$j$ and $y$ contribute to a weak H-bond in
Figure 2, for example, because the residue with
the lower number ($y$, since it is closer to the N-terminus), is a
bridge partner to $i$, the residue immediately following $j$.  For a
strand pair between the first and last strands of a protein, the
larger and smaller numbered residues must be switched, to account for
the fully circular nature of $\beta$-barrels.  Residues outside the TM
regions (core and headgroup) were not considered in our calculations
of pairwise contacts.

\paragraph*{Interstrand two-body spatial contact propensities.}
The interstrand pairwise propensity ({\sc Tmsip}) $P(X,Y)$ of residue
types $X$ and $Y$ for each of the three types of pairwise contacts is
given by:
\[
P(X,Y) = \frac{f(X,Y)}{\mathbb{E}[f'(X,Y)]},
\]
where $f(X,Y)$ is the observed frequency of $X$-$Y$ contacts of a
specific type in the TM regions (core and headgroup), and
$\mathbb{E}[f'(X,Y)]$ is the expected frequency of $X$-$Y$ contacts in
a null model.  In Tables 2 and 3,
$f(X,Y)$ is listed under ``Obs.'', and
$\mathbb{E}[f'(X,Y)]$ is listed under ``Exp.''

In order to calculate $\mathbb{E}[f'(X,Y)]$, we choose a null model in
which residues within each of the two adjacent strands in a strand
pair are permuted exhaustively and independently, and each permutation
occurs with equal probability.  In this null model, an $X$-$Y$ contact
forms if in a permuted strand pair two contacting residues happen to
be type $X$ and type $Y$.  $\mathbb{E}[f'(x,y)]$ is then the expected
number of $X$-$Y$ contacts over the entire dataset.

{\it Contacts between residues of the same type.\/} When $X$ is the
same as $Y$, the probability $\mathbb{P}_{X,X}(i) $
of $i=f'(X,X)$ number of $X$-$X$ contacts
in a strand pair follows a hypergeometric distribution:
\[
\mathbb{P}_{X,X}(i) 
 = \frac{\binom{x_1}{i}\binom{l-x_1}{x_2-i}}{\binom{l}{x_2}},
\]
where $x_1$ is the number of residues of type $X$ on the first strand,
$x_2$ is the number of residues of type $X$ on the second strand, and
$l$ is the length of the strand pair (the lengths of the two strands
must be equal).  This mimics the random selection of residues from
one strand to pair up with residues from the other strand.  

This is the same as picking $x_2$ balls from an urn of $l$ balls, $i$
of which are of type $X$ and $x_2-i$ of which are not type $X$.  In
this case, the urn represents the first strand, containing $x_1$
residues of type $X$ and $l-x_1$ residues not of type $X$.  The $x_2$
balls picked from the urn represent the residues in the first strand
selected to be placed adjacent to the $x_2$ residues of type $X$ in
the second strand, $i$ of which are of type $X$, and $x_2-i$ of which
are not of type $X$.

$\mathbb{E}_{\rm all}[f'(X,X)]$ is then the sum of the expected values
of $f'(X,X)$ for the set ${\cal SP}$ of all strand pairs in the
dataset.
\[
\mathbb{E}_{\rm all}[f'(X,X)] 
= \sum_{sp \in {\cal SP}} \mathbb{E}_{sp}[f'(X,X)]
= \sum_{sp \in {\cal SP}}
\frac{x_1(sp)\cdot x_2(sp)}{l(sp)},
\]
where $x_1(sp)$ and $x_2(sp)$ are the numbers of residues of type $X$
in the first and second strand of strand pair $sp \in {\cal SP}$,
respectively, and $l(sp)$ is the length of strand pair $sp$.  The
right-hand side is determined by using the expectation of the
hypergeometric distribution, analogous to Equation~(\ref{eqn:hyper2}).
For statistical significance, two-tailed $p$-values can be calculated
using formulae similar to Equations~(\ref{eqn:lowpvalue}) and
(\ref{eqn:highpvalue}).

{\it Contacts between residues of different types.\/} If the two
contacting residues are not of the same type, {\it i.e.}\ $X \neq Y$,
then the number of $X$-$Y$ contacts in the random model for one strand
pair is the sum of two dependent hypergeometric variables, one
variable for type $X$ residues in the first strand and type $Y$ in the
second strand, and another variable for type $Y$ residues in the first
strand and type $X$ in the second strand.  The expected frequency of
$X$-$Y$ contacts $\mathbb{E}[f'(X,Y)]$ is the sum of the two expected
values over all strand pairs $sp \in {\cal SP}$:
\[
\mathbb{E}[f'(X,Y)] 
= \sum_{sp \in {\cal SP}}\{
\mathbb{E}[f'_{sp}(X,Y)] +
\mathbb{E}[f'_{sp}(Y,X)]\}
= \sum_{sp \in {\cal SP}}\{
\frac{x_1(sp)\cdot y_2(sp)}{l(sp)} + \frac{y_1(sp)\cdot x_2(sp)}{l(sp)}
\},
\]
where $x_1(sp)$ and $x_2(sp)$ are the numbers of residues of type $X$
in the first and second strand, $y_1(sp)$ and $y_2(sp)$ are the
numbers of residues of type $Y$ in the first and second strand, and
$l(sp)$ is the length of strand pair $sp$.  The right-hand side is
determined by using the expectation of the hypergeometric
distribution, analogous to Equation~(\ref{eqn:hyper2}).  Despite the fact
that the variables $f'_{sp}(X,Y)$ and $f'_{sp}(Y,X)$ are dependent
({\it i.e.}\ the placement of an $X$-$Y$ pair may affect the
probability of a $Y$-$X$ pair in the same strand pair), their
expectations may be summed directly, because expectation is a linear
operator.

{\it Generalized hypergeometric model.}  However, because
$f'_{sp}(X,Y)$ and $f'_{sp}(Y,X)$ are dependent, to determine
$p$-values for a specific number of observed $X$-$Y$ contacts, a more
complex hypergeometric formula for the null model must be established.
The probability of a specific number of $X$-$Y$ contacts occurring in
one strand pair does not follow a simple hypergeometric distribution.
Here we develop a generalized hypergeometric model based on the
trinomial coefficient to characterize such a probability.  First, we
have a 3-element trinomial function $(a,b,c)!$ defined as:
\[
(a,b,c)! \equiv \frac{(a+b+c)!}{a!b!c!}.
\]
It represents the number of distinct permutations in a multiset of
three different types of elements, with number count $a, b$, and $c$
for each of the three element types.  Consider residues in the first
strand of length $l$ of a strand pair.  These $l$ residues are of
three types: $x_1$ count of type $X$ residues, $y_1$ of type $Y$
residues , and $l-x_1-y_1$ count of type ``neither''.  If we
exhaustively permute the $l$ residues, we have the trinomial
coefficient number of different permutations.  We denote this as:
\[
T(l, x_1, y_1) \equiv (x_1, y_1, l-x_1-y_1)!.
\]

We now first fix the positions of residues on strand 1, and permute
exhaustively all matching $l$ residues on strand 2.  Let $x_2, y_2$,
and $l-x_2-y_2$ be the numbers of residue of type $X$, $Y$, and
``neither'' on strand 2, respectively.  The total number of
permutations for strand 2 is:
\[
T(l, x_2, y_2) = (x_2, y_2, l-x_2-y_2)!.
\]

\begin{figure}[t!]
      \centerline{\epsfig{figure=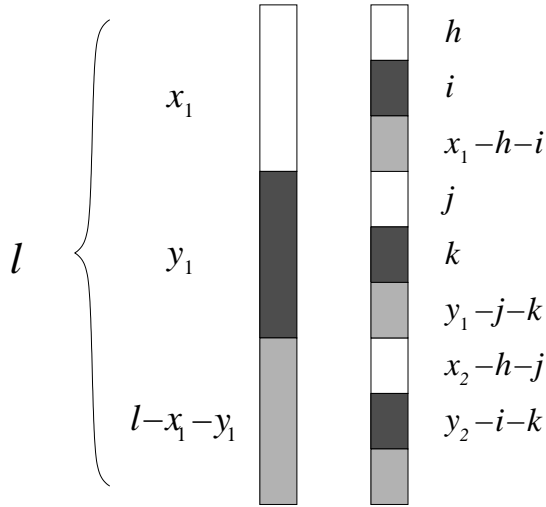,width=2.8in}}
\caption{\small \sf Illustration of the null model for
interstrand spatial motifs when $X \ne Y$.  White represents $X$
residues, black $Y$ residues, and grey ``neither'' residues ({\it
i.e.}\ neither X nor Y).  $X\mbox{-}Y$ motifs are represented by the
$i$ residue pairs in which there is an $X$ residue in the first strand
and a $Y$ residue in the second, and by the $j$ residue pairs in which
there is a $Y$ residue in the first strand and an $X$ residue in the
second.}
\label{Fig:math}
\end{figure}

Consider the residues on strand 2 that match to the $x_1$ number of
residues of type $X$ on strand 1. (This and all further descriptions
are illustrated in Figure~\ref{Fig:math} for clarification.)  These
$x_1$ residues on strand 2 consist of $h$ number of type $X$ residues,
$i$ number of type $Y$ residues, and $x_1 - h - i$ number of type
``neither'' residues.  They can be permuted in
\[
T(x_1, h, i) = (h, i, x_1 - h - i)!
\]
different ways.  By analogy, the $y_1$ residues on strand 2 that match
type $Y$ residues in strand 1 consist of $j$ number of type $X$
residues, $k$ number of type $Y$ residues, and $y_1 - j -k$ of type
``neither'' residues, and thus the total number of permutations for
these $y_1$ residues is:
\[
T(y_1, j, k) = (j, k, y_1 - j - k)!.
\]
Similarly, there are $T(l-x_1-y_1, \, x_2-h-j, \, y_2-i-k)$ number of
permutations to match the remaining $l-x_1-y_1$ of type ``neither''
residues on strand 1.

We characterize the probability $\mathbb{P}(h,i,j,k)$ of interstrand
matches: a) the $x_1$ type $X$ residues on strand 1 with $h$ type $X$
residues, $i$ type $Y$ residues, and $x_1 -h -i$ type ``neither''
residues on strand 2; b) the $y_1$ type $Y$ residues on strand 1 with
$j$ type $X$ residues, $k$ type $Y$ residues, and $y_1 -j -k$ type
``neither'' residues on strand 2; and c) the remaining $l-x_1-y_1$
type ``neither'' residues on strand 1 with $x_2 - h - j$ type $X$
residues, $y_2 - i - k$ type $Y$ residues, and the remaining type
``neither'' residues from strand 2. Equivalently,
$\mathbb{P}(h,i,j,k)$ is the probability of $h$ $X$-$X$ contacts, $i$
$X$-$Y$ contacts, $j$ $Y$-$X$ contacts, and $k$ $Y$-$Y$ contacts
occurring in a random permutation.

We introduce a higher order hypergeometric distribution for
$\mathbb{P}(h,i,j,k)$ as follows:
\[
\mathbb{P}(h,i,j,k) = \frac{T(x_1,h,i) \cdot T(y_1,j,k) \cdot T(
l-x_1-y_1, x_2-h-j, y_2-i-k)}{T(l,x_2,y_2)}.
\]
This can be illustrated as follows.  When randomly picking $x_2$ of
type $X$ residues, $y_2$ of type $Y$ residues, and the remaining
$l-x_2-y_2$ type ``neither'' residues from an urn for strand 2, we have:
(1) those matching the $x_1$ residues of type $X$ on strand 1 are of
$h$ number of type $X$, $i$ number of type $Y$, and $x_1 - h - i $
of type ``neither'';
(2) those matching the $y_1$ residues of type $Y$ on strand 1 are of
$j$ number of type $X$, $k$ number of type $Y$, and $x_2 - j - k $
of type ``neither''; and
(3) those matching the $l-x_1-y_1$ residues of type ``neither'' on
strand 1 are of $x_2 - h - j$ number of type $X$, $y_2 - i - k$ number
of type $Y$, and $(l_1 - x_1 -y_1) -(x_2 - h - j) - (y_2 - i -k)$ of
type ``neither''.

The marginal probability $\mathbb{P}_{X,Y}(m)$ that there are a total
of $i+j=m$ $X$-$Y$ contacts in the random model, namely, the pairings
where a residue of type $X$ in the first strand is paired with a
residue of type $Y$ in the second strand, summed with the pairings in
which a residue of type $Y$ in the first strand is paired with a
residue of type $X$ in the second strand is:
\[
\mathbb{P}_{X,Y}(m) = \sum_{h=0}^{x_1}\sum_{i=0}^{x_1-h}\sum_{k=0}^
{
\substack{
y_1-\\
(m-i)}
}
{
\mathbb{P}(h,i,m-i,k)},
\]
where $h$ is the number of matched $X$-$X$ contacts, $i$ the number of
matched $X$-$Y$ contacts, $m-i$ the number of matched $Y$-$X$
contacts($j$ in Figure~\ref{Fig:math}), and $k$ the number of matched
$Y$-$Y$ contacts.  The remaining contacts involving residues of type
``neither'' will then automatically be assigned, since all matches
involving $X$ and $Y$ have been accounted for.  There are $x_1$
possible values for $h$, one for each residue of type $X$ on strand 1;
$x_1-h$ possible values for $i$, once $h$ has been determined; and
$y_1-j=y-(m-i)$ possible values for $k$, once $i$ has been determined.
The $i$ number of $X$-$Y$ contacts plus the $m-i$ number of $Y$-$X$
contacts will sum to the $m$ number of contacts desired.  This
closed-form formula allows us to calculate analytically the two-tailed
$p$-value for this null model of $f'(X,Y)$ number of observed $X$-$Y$
contacts using formulae similar to Equations~(\ref{eqn:lowpvalue}) and
(\ref{eqn:highpvalue}).

{\it Confounding between single-body propensity and interstrand
two-body propensity.}  Because single-body propensities can vary
significantly, it is possible that differences in two-body
propensities may simply be reflections of differences in single-body
propensities, {\it e.g.}\ two polar residues might have high strong
H-bond pairwise propensities simply because both residues have an
independent preference for the same side of the TM barrel
(internal-facing), and not because of any direct significant
propensity between the two.  This artifact can be eliminated by
dividing each strand into two ``substrands,'' each of which contains
only residues facing the same direction.  This correction is automatic
for strong H-bonds and non-H-bonded interactions, as all of the
residues participating in each of these interactions in a single
strand pair face the same direction ({\it e.g.}\ the residues in a
particular strand pair participating in a strong H-bond must either
all be internal or all be external).  For weak H-bond interactions, in
which one residue is internal and one is external, each strand pair
must be divided into two substrand pairs: one pair in which the first
substrand is internal and the second is external, and another pair in
which the first substrand is external and the second is internal.  In
this way, the often dominating effects of single-residue orientation
are removed from two-body propensity calculation. Results reported in
Tables 2 and 3 are obtained after
these corrections.  For the analysis performed on the soluble
$\beta$-sheet dataset,
there is no strong distinction between internal and external residues,
since only some of the proteins are closed barrels.  Thus, no
correction was used for weak H-bonds.

\paragraph*{Pairwise propensities for a reduced alphabet.}
To obtain an objective reduced alphabet of amino acids for studying
membrane $\beta$-barrel proteins, we cluster amino acids by their
location preference and strand pairing propensities.  We represent
each amino acid as a vector and use hierarchical clustering to define
residue groups.  Each vector consists of 68 $z$-scores: one for each of
the 20 pairwise contact propensities including the residue for each of
the 3 interaction types (strong H-bonds, non-H-bonded interactions,
and weak H-bonds), and one for each single-body regional propensity
of the residue in each of the 8 regions.  The $z$-scores for pairwise
propensities are calculated as
\begin{equation}
z(X,Y) = \frac{f(X,Y)-\mathbb{E}[f'(X,Y)]}{\sqrt{\var{[f'(X,Y)]}}},
\label{eqn:zscore}
\end{equation}
and the $z$-scores for single-body propensities are calculated as
\[
z(X|r) = \frac{f(X|r)-\mathbb{E}[f'(X|r)]}{\sqrt{\var{[f'(X|r)]}}}.
\]

We use hierarchical clustering by average linkage with a Euclidean
distance function to obtain the clustering shown in
Figure 3.  We place the distance threshold so an
alphabet of 5 residues is formed.

\paragraph*{Strand register prediction.}
Our algorithm consists of two steps, the prediction of exact strand
starts and the prediction of strand register.  We use two sequence
models for these two tasks, namely, a 16-residue single strand model
and a 9-residue TM strand pair model (Figure~\ref{Fig:strandmodel}).
The regional designations in the 16-residue canonical strand model are
based on known physical attributes of TM $\beta$-strands: an average
length of 9-10 residues in the headgroup and core regions, and an
alternating internal-external pattern.
The size of each region (the cap regions, headgroup regions, and core
region) is determined by dividing the total number of residues in a
particular region by the number of strands in the dataset.  We have 4
residues for the extracellular cap region, 3 for the periplasmic cap
region, 2 each for the two headgroup regions, and 5 for the core
region.

The 9-residue strand pair model is derived from the canonical
16-residue models for two adjacent strands.  These 9 residues are
those designated to be in the transmembrane region, {\it i.e.}, in the
headgroup or core regions.  We exclude the 7 cap residues because the
cap regions do not contribute to strand pairing.  We also incorporate
the physical properties of the 3 types of strand interactions in the
model: The strong H-bond and non-H-bonded interactions must
alternate, and the weak H-bonds must extend one residue in the N-C
direction.  Due to chirality constraints in antiparallel
$\beta$-sheets ({\it i.e.}\ all amino acids in biological proteins are
L-amino acids), the backbone H-bonding pattern is fixed once the
N-C direction of the strands and the internal-external pattern are
determined by step 1.  We describe steps 1 and 2 in more detail:

\begin{figure}[t]
      \centerline{\epsfig{figure=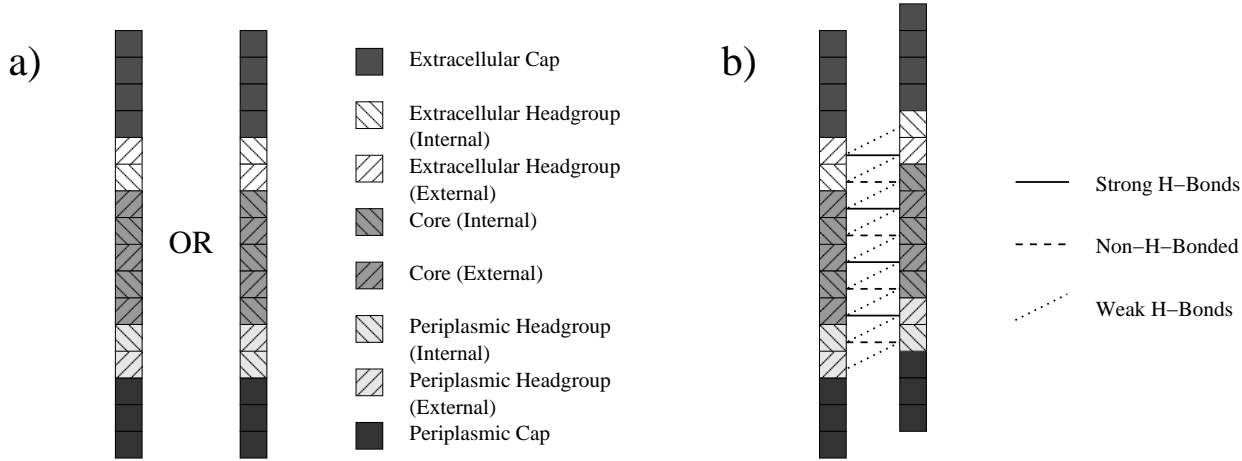,width=6.5in}}
\caption{\small \sf Illustrations of the models for
the strand register prediction algorithm.  a) 16-residue TM
$\beta$-strand model.  Two models are scored for each window with
different internal-external designations (hatch marks).  b) 9-residue
strand pair model.  The example shown has a strand shear of +1.  One
strand is shifted up or down against the other (fixed) strand to score
different strand shears. }
\label{Fig:strandmodel}
\end{figure}

\begin{itemize}
\item[1] {\it Predict exact starts.}  For a 16-residue window fitted
to our strand model (Figure~\ref{Fig:strandmodel}a), we calculate the
{\it single strand energy} $E(s;i)$ in $kT$ units for strand $i$:
\[
E(s;i) = - \sum_{k=1}^{16}\ln P(k;i,s),
\]
where $s$ is the displacement of the window position and $P(k;i,s)$ is
the single body propensity for the $k$-th residue in this window of
strand $i$, where the residues are alternatively assigned as internal
and external.  We calculate the energy score for a total of 11
possible windows: the window starting at the given approximated strand
start $s=0$ (either provided beforehand or calculated by a strand
predictor), and all windows 5 residues up ($s \le +5$) or down ($s \ge
-5$) in the amino acid sequence.  Because there are two possibilities
for the internal-external pattern of TM residues in our model, we
calculated strand energy scores for both possibilities for each
window.  We take the window with the lowest energy as the exact strand
start that we use in step 2.  The prediction also identifies which
residues are internal and which are external.

\item[2]{\it Predict strand register.}  After excluding the 7 cap
residues, we fit the 9-residue windows for two adjacent strands to the
strand pair model (Figure~\ref{Fig:strandmodel}b), and calculate the
{\it strand pairing energy} $E(s;i,i+1)$ in $kT$ units for the strand
pair $(i,i+1)$ with {\it strand shear} $s$ as
\[
\begin{split}
E(s;i,i+1) = &- \sum{\ln P_{SHB}(k;i,i+1,s)} - \sum{\ln P_{NB}(k;i,i+1,s)} \\
             &- \sum{\ln P_{WHB}(k;i,i+1,s)} + \alpha|\frac{N+2}{N} - s|.
\end{split}
\]
Here $i$ and $i+1$ are the labels of the two adjacent strands,
$P(k;i,i+1,s)$ is the propensity of pairing between the $k$-th
residues in strand $i$ and strand $i+1$, and the strand shear $s$ is
the residue displacement between the starts of the two strands.  The
first 3 terms refer to each of the 3 interaction types: strong H-bonds
(SHB), non-H-bonded interactions (NB), and weak H-bonds (WHB).  The
last term represents a penalty to the score when the strand shear $s$
deviates from the average strand shear, approximated as the average
shearing number $N+2$ of known TM $\beta$-barrels divided by the
number of strands $N$.  $\alpha$ is a coefficient determined
computationally.  We find that $\alpha=2$ works well.

We calculate the strand pairing score for a total of 11 possible
windows by sliding one strand in a pair against the other strand: the
windows with a strand shear of $s=-4$ residues up to a strand shear of
$s=+6$.  We also exclude strand shears that place internal residues
next to external residues, as this would violate the H-bonding
patterns of $\beta$-sheets.  This reduces the search space by half.
The strand shear $s$ with the lowest strand pairing energy is taken as
the true strand register.
\end{itemize}

To approximate the strand starts of the $\beta$-barrel membrane
proteins in our dataset, we use the hidden Markov model predictor of
Bigelow {\it et al.}\ \cite{Bigelow04_NAR}, one of the most successful
predictors for strand starts.  This predictor uses only the amino acid
sequence as input, and outputs a designation for each residue from a
list of 4 possibilities: ``up-strand'' (referring to an odd-numbered
strand), ``down-strand'' (even-numbered strand), loop between an up-
and down-strand, and loop between a down- and up-strand.  This
predictor will therefore also predict the number of strands in the
protein {\it ab initio}.  We use the first instances of the up- and
down-strand designations for each strand as approximate strand starts.

We exclude from our prediction analysis two proteins in our dataset (6
strand pairs) for which the strand start predictor failed: TolC and
$\alpha$-HL.  For TolC, the predictor did not detect any TM strands.
For $\alpha$-HL, the predictor detected far too many strands (8
instead of the actual 2 strands).  For the remaining 17 of the 19
proteins, the hidden Markov model correctly predicted 252 of the 256
strands, with only 4 false positives.  The false positive strands were
all short (7 residues) and had irregular composition.  Our register
prediction was also incorrect for all strand pairs involving these 4
false positives.  We apply our prediction to the 17 $\beta$-barrel
membrane proteins in leave-one-out fashion: we calculate single and
pairwise propensities using 16 of the proteins, and use them to
predict strand registers in the 17th protein.  Since the pairwise
propensities are derived from a very small dataset, we introduce a
pseudocount of 1 to the observed and expected numbers of each pair
when calculating propensities.


\begin{thebibliography}{10}
\expandafter\ifx\csname url\endcsname\relax
  \def\url#1{\texttt{#1}}\fi
\expandafter\ifx\csname urlprefix\endcsname\relax\def\urlprefix{URL }\fi

\bibitem{Montoya03_BBA}
M.~Montoya, E.~Gouaux, {$\beta$-barrel membrane protein folding and structure
  viewed through the lens of $\alpha$-hemolysin}, Biochim Biophys Acta. 1609
  (2003) 19--27.

\bibitem{Tamm01_JBC}
L.~K. Tamm, A.~Arora, J.~H. Kleinschmidt, {Structure and assembly of
  $\beta$-barrel membrane proteins}, J Biol Chem. 276 (2001) 32399--32402.

\bibitem{Wimley03_COSB}
W.~C. Wimley, {The versatile $\beta$-barrel membrane protein}, Curr Opin Struct
  Biol. 13 (2003) 404--411.

\bibitem{Song96_Sci}
L.~Song, M.~R. Hobaugh, C.~Shustak, S.~Cheley, H.~Bayley, J.~E. Gouaux,
  {Structure of staphylococcal $\alpha$-hemolysin, a heptameric transmembrane
  pore}, Science. 274 (1996) 1859--1866.

\bibitem{Nassi02_BC}
S.~Nassi, R.~J. Collier, A.~Finkelstein, {PA$_{63}$ channel of anthrax toxin:
  an extended $\beta$-barrel}, Biochemistry. 41 (2002) 1445--1450.

\bibitem{Koebnik00_MM}
R.~Koebnik, K.~P. Locher, P.~Van~Gelder, {Structure and function of bacterial
  outer membrane proteins: barrels in a nutshell}, Mol Microbiol. 37 (2000)
  239--253.

\bibitem{Bigelow04_NAR}
H.~R. Bigelow, D.~S. Petrey, J.~Liu, D.~Przybylski, B.~Rost, {Predicting
  transmembrane $\beta$-barrels in proteomes}, Nucleic Acids Res. 32 (2004)
  2566--2577.

\bibitem{Martelli02_Bioinfo}
P.~L. Martelli, P.~Fariselli, A.~Krogh, R.~Casadio, {A sequence-profile-based
  HMM for predicting and discriminating $\beta$ barrel membrane proteins},
  Bioinformatics. 18 Suppl 1 (2002) S46--S53.

\bibitem{Gromiha04_JCC}
M.~M. Gromiha, S.~Ahmad, M.~Suwa, {Neural network-based prediction of
  transmembrane $\beta$-strand segments in outer membrane proteins}, J Comput
  Chem. 25 (2004) 762--767.

\bibitem{Schulz00_COSB}
G.~E. Schulz, {$\beta$-barrel membrane proteins}, Curr Opin Struct Biol. 10
  (2000) 443--447.

\bibitem{Schulz02_BBA}
G.~E. Schulz, {The structure of bacterial outer membrane proteins}, Biochim
  Biophys Acta. 1565 (2002) 308--317.

\bibitem{Tamm04_BBA}
L.~K. Tamm, H.~Hong, B.~Liang, {Folding and assembly of $\beta$-barrel membrane
  proteins}, Biochim Biophys Acta. 1666 (2004) 250--263.

\bibitem{Seshadri98_PS}
K.~Seshadri, R.~Garemyr, E.~Wallin, G.~von Heijne, A.~Elofsson, {Architecture
  of $\beta$-barrel membrane proteins: analysis of trimeric porins}, Protein
  Sci. 7 (1998) 2026--2032.

\bibitem{Wimley02_PS}
W.~C. Wimley, {Toward genomic identification of $\beta$-barrel membrane
  proteins: composition and architecture of known structures}, Protein Sci. 11
  (2002) 301--312.

\bibitem{Ulmschneider01_BBA}
M.~B. Ulmschneider, M.~S. Sansom, {Amino acid distributions in integral
  membrane protein structures}, Biochim Biophys Acta. 1512 (2001) 1--14.

\bibitem{Gromiha03_IJBM}
M.~M. Gromiha, M.~Suwa, {Variation of amino acid properties in all-$\beta$
  globular and outer membrane protein structures}, Int J Biol Macromol. 32
  (2003) 93--98.

\bibitem{Chamberlain04_PS}
A.~K. Chamberlain, J.~U. Bowie, {Asymmetric amino acid compositions of
  transmembrane $\beta$-strands}, Protein Sci. 13 (2004) 2270--2274.

\bibitem{Wouters95_Prot}
M.~A. Wouters, P.~M. Curmi, {An analysis of side chain interactions and pair
  correlations within antiparallel $\beta$-sheets: the differences between
  backbone hydrogen-bonded and non-hydrogen-bonded residue pairs}, Proteins. 22
  (1995) 119--131.

\bibitem{Steward02_Prot}
R.~E. Steward, J.~M. Thornton, {Prediction of strand pairing in antiparallel
  and parallel $\beta$-sheets using information theory}, Proteins. 48 (2002)
  178--191.

\bibitem{Merkel98_FD}
J.~S. Merkel, L.~Regan, {Aromatic rescue of glycine in $\beta$ sheets}, Fold
  Des. 3 (1998) 449--455.

\bibitem{Hutchinson98_PS}
E.~G. Hutchinson, R.~B. Sessions, J.~M. Thornton, D.~N. Woolfson, {Determinants
  of strand register in antiparallel $\beta$-sheets of proteins}, Protein Sci.
  7 (1998) 2287--2300.

\bibitem{Granseth05_JMB}
E.~Granseth, G.~von Heijne, A.~Elofsson, {A study of the membrane-water
  interface region of membrane proteins.}, J Mol Biol. 346 (2005) 377--385.

\bibitem{Adamian01_JMB}
L.~Adamian, J.~Liang, {Helix-helix packing and interfacial pairwise
  interactions of residues in membrane proteins}, J Mol Biol. 311 (2001)
  891--907.

\bibitem{vonHeijne89_Nat}
G.~von Heijne, {Control of topology and mode of assembly of a polytopic
  membrane protein by positively charged residues}, Nature. 341 (1989)
  456--458.

\bibitem{Chamberlain04_JMB}
A.~K. Chamberlain, Y.~Lee, S.~Kim, J.~U. Bowie, {Snorkeling preferences foster
  an amino acid composition bias in transmembrane helices}, J Mol Biol. 339
  (2004) 471--479.

\bibitem{Wilmot88_JMB}
C.~M. Wilmot, J.~M. Thornton, {Analysis and prediction of the different types
  of $\beta$-turn in proteins}, J Mol Biol. 203 (1988) 221--232.

\bibitem{Schulz04_BAEP}
G.~E. Schulz, {The structures of general porins}, in: R.~Benz (Ed.), Bacterial
  and Eukaryotic Porins, WILEY-VCH, 2004, pp. 25--40.

\bibitem{Adamian02_Prot}
L.~Adamian, J.~Liang, {Interhelical hydrogen bonds and spatial motifs in
  membrane proteins: polar clamps and serine zippers}, Proteins. 47 (2002)
  209--218.

\bibitem{Adamian03_JMB}
L.~Adamian, R.~Jackups, T.~A. Binkowski, J.~Liang, {Higher-order interhelical
  spatial interactions in membrane proteins}, J Mol Biol. 327 (2003) 251--272.

\bibitem{Eilers02_BJ}
M.~Eilers, A.~B. Patel, W.~Liu, S.~O. Smith, {Comparison of helix interactions
  in membrane and soluble $\alpha$-bundle proteins}, Biophys J. 82 (2002)
  2720--2736.

\bibitem{Liu04_JMB}
W.~Liu, M.~Eilers, A.~B. Patel, S.~O. Smith, {Helix packing moments reveal
  diversity and conservation in membrane protein structure}, J Mol Biol. 337
  (2004) 713--729.

\bibitem{Gimpelev04_BJ}
M.~Gimpelev, L.~R. Forrest, D.~Murray, B.~Honig, {Helical packing patterns in
  membrane and soluble proteins}, Biophys J. 87 (2004) 4075--4086.

\bibitem{Ho02_JMB}
B.~K. Ho, P.~M. Curmi, {Twist and shear in $\beta$-sheets and $\beta$-ribbons},
  J Mol Biol. 317 (2002) 291--308.

\bibitem{Senes01_PNAS}
A.~Senes, I.~Ubarretxena-Belandia, D.~M. Engelman, {The C$\alpha$--H$\cdots$O
  hydrogen bond: a determinant of stability and specificity in transmembrane
  helix interactions}, Proc Natl Acad Sci U S A. 98 (2001) 9056--9061.

\bibitem{Riddle97_NSB}
D.~S. Riddle, J.~V. Santiago, S.~T. Bray-Hall, N.~Doshi, V.~P. Grantcharova,
  Q.~Yi, D.~Baker, {Functional rapidly folding proteins from simplified amino
  acid sequences}, Nat Struct Biol. 4 (1997) 805--809.

\bibitem{Murphy00_PE}
L.~R. Murphy, A.~Wallqvist, R.~M. Levy, {Simplified amino acid alphabets for
  protein fold recognition and implications for folding}, Protein Eng. 13
  (2000) 149--152.

\bibitem{Li03_Prot}
X.~Li, C.~Hu, J.~Liang, {Simplicial edge representation of protein structures
  and alpha contact potential with confidence measure}, Proteins. 53 (2003)
  792--805.

\bibitem{Shindyalov98_PE}
I.~N. Shindyalov, P.~E. Bourne, {Protein structure alignment by incremental
  combinatorial extension (CE) of the optimal path.}, Protein Eng. 9 (1998)
  739--747.

\bibitem{White99_ARBBS}
S.~H. White, W.~C. Wimley, {Membrane protein folding and stability: physical
  principles}, Annu Rev Biophys Biomol Struct. 28 (1999) 319--365.

\bibitem{Chamberlain04_BJ}
A.~K. Chamberlain, J.~U. Bowie, {Analysis of side-chain rotamers in
  transmembrane proteins}, Biophys J. 87 (2004) 3460--3469.

\bibitem{Nakae86_CRM}
T.~Nakae, {Outer-membrane permeability of bacteria}, Crit Rev Microbiol. 13
  (1986) 1--62.

\bibitem{deCock96_EMBO}
H.~de~Cock, J.~Tommassen, {Lipopolysaccharides and divalent cations are
  involved in the formation of an assembly-competent intermediate of
  outer-membrane protein PhoE of {\it E. coli}}, EMBO J. 15 (1996) 5567--5573.

\bibitem{Ferguson98_Sci}
A.~D. Ferguson, E.~Hofmann, J.~W. Coulton, K.~Diederichs, W.~Welte,
  {Siderophore-mediated iron transport: crystal structure of FhuA with bound
  lipopolysaccharide}, Science. 282 (1998) 2215--2220.

\bibitem{Nikaido03_MMBR}
H.~Nikaido, {Molecular basis of bacterial outer membrane permeability
  revisited}, Microbiol Mol Biol Rev. 67 (2003) 593--656.

\bibitem{Ramakrishnan04_BC}
M.~Ramakrishnan, C.~L. Pocanschi, J.~H. Kleinschmidt, D.~Marsh, {Association of
  spin-labeled lipids with $\beta$-barrel proteins from the outer membrane of
  {\it Escherichia coli}}, Biochemistry. 43 (2004) 11630--11636.

\bibitem{vanKlompenburg97_EMBO}
W.~van Klompenburg, I.~Nilsson, G.~von Heijne, B.~de~Kruijff, {Anionic
  phospholipids are determinants of membrane protein topology}, EMBO J. 16
  (1997) 4261--4266.

\bibitem{White04_COSB}
S.~H. White, G.~von Heijne, {The machinery of membrane protein assembly}, Curr
  Opin Struct Biol. 14 (2004) 397--404.

\bibitem{Kleinschmidt99_Biochem}
J.~H. Kleinschmidt, L.~K. Tamm, {Time-resolved distance determination by
  tryptophan fluorescence quenching: probing intermediates in membrane protein
  folding}, Biochemistry. 38 (1999) 4996--5005.

\bibitem{Weikl03_JMB}
T.~R. Weikl, K.~A. Dill, {Folding kinetics of two-state proteins: effect of
  circularization, permutation, and crosslinks.}, J Mol Biol. 332 (2003)
  953--963.

\bibitem{Kabsch83_Biopoly}
W.~Kabsch, C.~Sander, {Dictionary of protein secondary structure: pattern
  recognition of hydrogen-bonded and geometrical features}, Biopolymers. 22
  (1983) 2577--2637.

\bibitem{Pautsch98_NSB}
A.~Pautsch, G.~E. Schulz, {Structure of the outer membrane protein A
  transmembrane domain}, Nat Struct Biol. 5 (1998) 1013--1017.

\bibitem{Vogt99_SFD}
J.~Vogt, G.~E. Schulz, {The structure of the outer membrane protein OmpX from
  {\it Escherichia coli} reveals possible mechanisms of virulence}, Structure
  Fold Des. 7 (1999) 1301--1309.

\bibitem{Vandeputte03_JBC}
L.~Vandeputte-Rutten, M.~P. Bos, J.~Tommassen, P.~Gros, {Crystal structure of
  Neisserial surface protein A (NspA), a conserved outer membrane protein with
  vaccine potential}, J Biol Chem. 278 (2003) 24825--24830.

\bibitem{Prince02_PNAS}
S.~M. Prince, M.~Achtman, J.~P. Derrick, {Crystal structure of the OpcA
  integral membrane adhesin from {\it Neisseria meningitidis}}, Proc Natl Acad
  Sci U S A. 99 (2002) 3417--3421.

\bibitem{Vandeputte01_EMBO}
L.~Vandeputte-Rutten, R.~A. Kramer, J.~Kroon, N.~Dekker, M.~R. Egmond, P.~Gros,
  {Crystal structure of the outer membrane protease OmpT from {\it Escherichia
  coli} suggests a novel catalytic site}, EMBO J. 20 (2001) 5033--5039.

\bibitem{Snijder99_Nat}
H.~J. Snijder, I.~Ubarretxena-Belandia, M.~Blaauw, K.~H. Kalk, H.~M. Verheij,
  M.~R. Egmond, N.~Dekker, B.~W. Dijkstra, {Structural evidence for
  dimerization-regulated activation of an integral membrane phospholipase},
  Nature. 401 (1999) 717--721.

\bibitem{Oomen04_EMBO}
C.~J. Oomen, P.~Van~Ulsen, P.~Van~Gelder, M.~Feijen, J.~Tommassen, P.~Gros,
  {Structure of the translocator domain of a bacterial autotransporter}, EMBO
  J. 23 (2004) 1257--1266.

\bibitem{Weiss92_JMB}
M.~S. Weiss, G.~E. Schulz, {Structure of porin refined at 1.8 \AA\ resolution},
  J Mol Biol. 227 (1992) 493--509.

\bibitem{Kreusch94_JMB}
A.~Kreusch, G.~E. Schulz, {Refined structure of the porin from {\it
  Rhodopseudomonas blastica}. Comparison with the porin from {\it Rhodobacter
  capsulatus}}, J Mol Biol. 243 (1994) 891--905.

\bibitem{Cowan95_Struct}
S.~Cowan, R.~M. Garavito, J.~N. Jansonius, J.~A. Jenkins, R.~Karlsson,
  N.~Konig, E.~F. Pai, R.~A. Pauptit, P.~J. Rizkallah, J.~P. Rosenbusch,
  et~al., {The structure of OmpF porin in a tetragonal crystal form},
  Structure. 3 (1995) 1041--1050.

\bibitem{Zeth00_SFD}
K.~Zeth, K.~Diederichs, W.~Welte, H.~Engelhardt, {Crystal structure of Omp32,
  the anion-selective porin from {\it Comamonas acidovorans}, in complex with a
  periplasmic peptide at 2.1 \AA\ resolution}, Structure Fold Des. 8 (2000)
  981--992.

\bibitem{Meyer97_JMB}
J.~E. Meyer, M.~Hofnung, G.~E. Schulz, {Structure of maltoporin from {\it
  Salmonella typhimurium} ligated with a nitrophenyl-maltotrioside}, J Mol
  Biol. 266 (1997) 761--775.

\bibitem{Forst98_NSB}
D.~Forst, W.~Welte, T.~Wacker, K.~Diederichs, {Structure of the
  sucrose-specific porin ScrY from {\it Salmonella typhimurium} and its complex
  with sucrose}, Nat Struct Biol. 5 (1998) 37--46.

\bibitem{Buchanan99_NSB}
S.~K. Buchanan, B.~S. Smith, L.~Venkatramani, D.~Xia, L.~Esser, M.~Palnitkar,
  R.~Chakraborty, D.~van~der Helm, J.~Deisenhofer, {Crystal structure of the
  outer membrane active transporter FepA from {\it Escherichia coli}}, Nat
  Struct Biol. 6 (1999) 56--63.

\bibitem{Ferguson02_Sci}
A.~D. Ferguson, R.~Chakraborty, B.~S. Smith, L.~Esser, D.~Van Der~Helm,
  J.~Deisenhofer, {Structural basis of gating by the outer membrane transporter
  FecA}, Science. 295 (2002) 1715--1719.

\bibitem{Chimento03_NSB}
D.~P. Chimento, A.~K. Mohanty, R.~J. Kadner, M.~C. Wiener, {Substrate-induced
  transmembrane signaling in the cobalamin transporter BtuB}, Nat Struct Biol.
  10 (2003) 394--401.

\bibitem{Koronakis00_Nat}
V.~Koronakis, A.~J. Sharff, E.~Koronakis, B.~Luisi, C.~Hughes, {Crystal
  structure of the bacterial membrane protein TolC central to multidrug efflux
  and protein export}, Nature. 405 (2000) 914--919.

\end{thebibliography}
\end{document}